\documentclass[useAMS,usenatbib]{mn2e}
\usepackage{pdflscape}

\usepackage{epsfig,rotate,graphicx}
\usepackage{subfigure}

\usepackage{color}
\usepackage{ulem}
\usepackage{graphicx}
\usepackage{amsfonts}
\usepackage{amsmath}
\usepackage{amssymb}
\usepackage{psfrag}
\usepackage{combinedgraphics}
\usepackage[retainorgcmds]{IEEEtrantools}
\usepackage[cleanup={.log,.ps,.dvi,.bbl}]{pstool}
\usepackage{expl3}

\usepackage{txfonts}

\newcommand\bb[1]{\bf{#1}}
\newcommand\del{\nabla}
\newcommand\bcdot{\bb{\cdot}}
\newcommand\btimes{\bb{\times}}
\newcommand{\bmth}[1]{\mbox{\boldmath${#1}$}}
\DeclareMathOperator{\Tr}{Tr} 
\DeclareMathOperator{\ii}{i} 

\title[Stability of vortices]{On the local stability of vortices in differentially rotating discs}

\author[Railton and Papaloizou]{A. D. Railton $^1$\thanks{E-mail:
ar488@damtp.cam.ac.uk} and J. C. B. Papaloizou $^{1}$\thanks{E-mail:
jcbp2@damtp.cam.ac.uk} \\
$^1$ Department of Applied Mathematics and Theoretical Physics,\\
University of Cambridge, Centre for Mathematical Sciences,\\
Wilberforce Road, Cambridge, CB3 0WA, UK \\}

\begin{document}

\maketitle

\begin{abstract}
	In order to circumvent the loss of solid material through radial drift towards the central star, the trapping  of dust inside persistent vortices in protoplanetary discs has often been suggested as a process that can  eventually lead to planetesimal formation. Although a few special cases have been discussed, exhaustive studies of possible quasi-steady configurations available for dust-laden vortices and their stability have yet to be undertaken, thus their viability or otherwise as locations for the gravitational instability to take hold and seed planet formation is unclear. In this paper we generalise and extend the well known Kida solution to obtain a series of steady state solutions with varying vorticity and dust density distributions in their cores, in the limit of perfectly coupled dust and gas. We then present a local stability analysis of these configurations, considering perturbations localised on streamlines. Typical parametric instabilities found have growthrates of $~0.05\Omega_P$, where $\Omega_P$ is the angular velocity at the centre of the vortex. Models with density excess can exhibit many narrow parametric instability bands while those with a concentrated vorticity source display internal shear which significantly affects their stability. However, the existence of these parametric instabilities may not necessarily prevent the possibility of dust accumulation in vortices. 
\end{abstract}


\section{Introduction} \label{sec:intro}
	Studies of planet formation have been ongoing since the formulation of the nebula hypothesis for the formation and early evolution of the Solar System \citep{Swedenborg1734,Kant1755, Weizsaecker1944}. 
	Centrifugal forces for the most part balance the gravitational attraction of the central star and a protoplanetary (PP) disc that forms along with it. The disc contains dust grains $0.1-1 \mu \rm{m}$ in size which undergo coagulation \citep[eg.][]{Safronov1969, Dominik1997} through the action of electrostatic rather than gravitational forces, the latter being expected to dominate during the later stages of planet formation \citep[eg.][]{Safronov1969, Lissauer1993, Papaloizou2006}. \\

	The notion of planetesimal formation, through the sticking together of dust grains in PP discs through two-body collisions, was first developed by Chamberlin \citep[eg.][]{Chamberlin1900}. 
	However, bodies above about a meter in size have very poor sticking qualities \citep{Benz2000} so that collisions between them will result in fragmentation or bouncing rather than growth. 
	In addition, particles in a typical PP disc, with mid plane pressure decreasing monotonically with radius, experience a headwind in the azimuthal direction which causes them to lose angular momentum and drift radially towards the central star. As a consequence, metre-sized bodies may spiral into the star on timescales as short as a hundred years \citep[eg.][]{Weidenschilling1977, Papaloizou2006}. This indicates that planetesimals must be formed within the rapid radial drift time of these bodies. These two difficulties for planetesimal formation constitute the metre-size barrier.


	However, in a disc for which the mid plane pressure does not decrease monotonically with radius, aerodynamic effects can concentrate solids in allowing planetesimals to form.
	As a consequence of the fact that particles tend to drift in the direction of the pressure gradient, \citet{Whipple1972} showed that a pressure maximum located in an axisymmetric ring is a very effective particle trap. If sufficient concentration can occur, planetesimal formation can then be assisted by gravitational instability \citep{Safronov1969, Goldreich1973}. In the context of the above scenario, MRI simulations indicate the possibility of zonal flows that produce long-lived axisymmetric pressure bumps \citep{Johansen2009,Fromang2009}.

	Isolated pressure maxima can exist in the centre of anticyclonic vortices. \citet{Barge1995} showed that such vortices could also be natural localised particle traps, proposing that they could be sites of planetesimal formation. A number of authors have shown that anticyclonic vortices can form coherent and long-lived structures in PP discs \citep{Bracco1999,Chavanis2000,Barranco2005}.
	In addition there are diverse means of generating these vortices in discs, namely through the Baroclinic Instability \citep{Petersen2007, Lesur2010}, 
	the instability at the interface between MRI active and dead zones \citep{Lovelace1999,Meheut2010}, and the edge instability associated with gaps in the disc produced by existing planets \citep{Lin2011}.

	However, it is well known that such vortices are prone to the so called elliptical instability, which can be regarded as a local parametric instability associated with periodic motion on streamlines \citep[see][and references therein]{Lesur2009}. However, this has only been analysed in full detail for the special case of a Kida vortex \citep[see][]{Kida1981} with no dust present. These solutions apply to a local patch of the disc that can be represented using the well known shearing box formalism \citep{Goldreich1965}. However, as discussed in this paper, a large variety of vortex configurations can be constructed with and without dust concentrations. Their vorticity and density profiles have a degree of arbitrariness when no frictional or diffusive processes operate, although they would be expected to be determined by the form of these when they do.

	Since the existence of instabilities in such vortices could be a threat to their survival or dust attracting capability, a comprehensive stability analysis is desirable. However, up to now only velocity profiles that give a constant period of circulation such as in the Kida vortex has been considered in theoretical developments. In this context \citet{Chang2010} considered the 2D (independent of the vertical direction) stability of a dust laden vortex assuming such a profile but did not consider the issue of whether the profiles adopted provided either a steady state or matched onto a suitable disc background. They assumed a separability that applies strictly to the uniform density Kida solution but not more general cases. It is the purpose of this paper to further consider the structure and stability of vortices in a protoplanetary disc background.

	For simplicity we consider vortices with radial length scale less than the disc scale height for which the dust stopping time is very short. In this case an incompressible fluid model with frozen-in density distribution can be adopted. We consider vortices allowing for both non-uniform vorticity and density distributions in their cores. We consider local stability but with an emphasis of keeping the analysis as general as possible so that it is not necessary to have particular velocity profiles and we can address some of the issues mentioned above. Apart from recapturing the existing results for the Kida vortex we are also able to consider vortices with more general vorticity and density distributions and make an assessment of instabilities on the dust accumulation process. 

	The plan of the paper is as follows: In Section \ref{sec:basic_equations} we give the basic equations governing the fluid model that we use. We go on to derive a partial differential equation for the stream function in Section \ref{sec:steady_state_equations}. In Section \ref{polyeq} we adapt the Kida solution to apply to the situation when the vortex has a high density core. This polytropic solution is applicable to a Keplerian background when the aspect ratio is $7$. For other values the background has a superposed pressure extremum. 

	In Section \ref{stability} we formulate the stability analysis governing local perturbations to incompressible vortical flows. Perturbations localised on streamlines are considered. 
In an Eulerian description, these can be associated with a time-independent  wavenumber  that can lead to exponentially growing modes or in the generic case, where the period of circulation in the vortex is not constant
 a wavenumber with a magnitude  that ultimately increases linearly with time. The former class of modes is shown to give rise to the known instabilites of the Kida vortex.
 We generalise the instability seen there that is associated with a central saddle point in the pressure distribution to more general cases and indicate how parametric instability bands appear for streamlines close to the centre.
 Modes with wavenumbers whose magnitude  ultimately increases  linearly with time cannot have amplitudes that
increase exponentially with time indefinitely, but they may undergo temporary amplification. In the special case of a vortex with constant period, 
as was considered by \citet{Chang2010}, the corresponding
amplitudes may grow exponentially  with time.  

	We go on to describe our numerical procedures and results in Section \ref{sec:numerical_method}. We give results for steady state vortices with non-uniform vorticity sources in their cores, with and without increased central density on account of a dust component, and discuss their stability. We augment the discussion of stability by considering different modes for the analytic polytropic solution and also a simple point vortex model which should be a limiting case applicable to streamlines distant from the core.
	These models are found to behave according to expectation from the other numerical models.  Often narrow instability bands are found that incoming dust would have to encounter.

	In Section \ref{sec:conclusions} we summarise and discuss our results, providing arguments why instabilities of the type found here may not prevent significant dust accumulation in vortices with large aspect ratio. 
\section{Model and basic equations} \label{sec:basic_equations}
	\begin{table}
	\caption{Table of parameters, variables and symbols} \label{symbols}
        \vspace{-0.3cm}
 	\begin{center}
		\begin{tabular}{c p{6.2cm}}
		\hline
		Symbol & Definition \\
		\hline
		$\Omega_P $ & Magnitude of angular velocity  \bmth{\Omega_P} \\
		${\bf {\hat k} }$& Unit vector in $z$ direction\\
		$\rho,$ $\rho'$& Density and density perturbation\\
		${\bf v} \equiv (v_1,v_2,v_3)$ & Velocity\\
                ${\bf v}'\equiv (v_x', v_y', v_z')$& Velocity perturbation\\
		$P,$ $P'$& Pressure  and pressure perturbation \\
		$\Phi= \Phi_{gr}+ \Phi_{rot}$ & Sum of gravitational and centrifugal potentials \\
        ${\bf r}$& Position vector measured from the cental star of mass $M,$ $r= |{\bf r}|$\\
        $G$& Gravitational constant \\ 
		$\tau_s$ & Dust stopping time \\
		$c_s= H\Omega_P$& Sound speed with $H$ being the disc scale height\\
		$\psi, \psi_0, \psi_1 $ & Stream functions for the general flow, the  background flow and for the superposed vortex, $\psi=\psi_0+\psi_1$ \\
		$F(\psi)$ & Arbitrary function that appears in Poisson form of momentum equation, see equation~(\ref{eq:fq}) \\   
		$\mathcal{A}(\psi)$ & Bernoulli source term in equation~(\ref{eq:psi1}) \\
		$\mathcal{B}(\psi)$ & Density source term in equation~(\ref{eq:psi1}) \\
		$\alpha,\,\beta$ & Power-law indices in expressions for $\mathcal{A}(\psi),\,\mathcal{B}(\psi)$, respectively \\
		$A,\,B$ & Scaling factors in expressions for  $\mathcal{A}(\psi),\,\mathcal{B}(\psi)$, respectively \\
		$\chi$ & Aspect ratio of vortex patch \\
		$\psi_b$ & The value of $\psi$ evaluated on vortex boundary \\
		$S$ & Shear of background flow \\ 
		$\omega_t$ & Total vorticity in  vortex patch \\
		$\omega_v$ & Vorticity imposed on background to produce vortex patch \\
		$b$ & Parameter determining  magnitude of  central density excess (polytropic model) \\
		$n_1$ & Power-law index  in expression for density  profile (polytropic model) \\
		$\beta_P$ &  Scaling factor in  expression for pressure  (polytropic model) \\ 
		$\bmth{\xi}$& Lagrangian displacement \\
		$S_A$ & Phase function in WKBJ ansatz (local analysis) \\
		$\lambda$ & Large parameter in WKBJ ansatz (local analysis) \\
	    ${\bf k} \equiv (k_x,k_y,k_z)$& Wave vector\\
		$\theta, k_0, {\bar t}$ & Constant angle,  wavenumber scaling parameter \& constant of integration occurring in equation~(\ref{solution_k}) for $\bmth{k}$ \\
		$\sigma$& Eigenfrequency \\
		${\sf E}$ & Symmetric matrix on  the  RHS of equation~(\ref{2sem2}) \\
		$S_{\perp}$ & Part of $S_A$ that is a function of $x$ and $y$ only\\
		$C_K$ & Amplitude factor and constant of integration  in equation~(\ref{eq:part_traj})  for particle trajectories in Kida vortices \\
		$S_K, \phi_0$ & Constant factor and angle  in  expression for $S_{\perp}$ applicable to a  Kida vortex \\
		$\eta$ & Scaling  amplitude for the Eulerian  velocity perturbation in  vertical stability analysis\\
		$\tau,\,W$ & Scaled  time and $\eta$ in vertical stability analysis \\
		$\mathcal{Z}$& The  quantity $\propto D\rho'/Dt$ used  in the  vertical stability analysis which  satisfies a Hill  equation \\
		$q$ & Constant in Hill equation for $\mathcal{Z}$ \\
		$\omega_m$ & Measure of total imposed vorticity for numerical vortex models  \\
		$\rho_m$ & Measure of total imposed mass excess for numerical vortex models \\
		$\gamma$ & Growth rate of instability \\
		$\gamma_{_2}$ & Parameter taking  asymptotic form $\gamma_{_2} \to \gamma t^2/2$ used to estimate growth rate\\
		$\tilde{P}=2\pi/\omega,\,\tilde{P}_{\text{\sc{kida}}}$ & Period to circulate around  a streamline in the general and Kida cases \\
		\hline
		\end{tabular}
 	\end{center}
	\end{table}

	We begin by considering a single fluid model of the dust and gas circulating in a protoplanetary accretion disc. We consider unmagnetized regions of the disc such as dead zones and so neglect Lorentz forces. The basic equations for the fluid are those of continuity and momentum conservation. In a frame rotating with angular velocity ${\bb{\Omega}_P} = \Omega_P {\hat {\bf k}},$ with ${\hat {\bf k}}$ being the unit vector in the fixed direction of rotation (here called the vertical direction) and $\Omega_P$ being the magnitude of the angular velocity, these take the form 
	\begin{equation}
		{{\partial \rho}\over {\partial t}} + \del \bcdot( \rho {\bb{v}} ) = 0 \label{eq.cont} 
	\end{equation}
	and
	\begin{equation}
		 \rho\left(\frac{\partial {\bb{v}}}{\partial t} + {\bb{v}} \bcdot \del {\bb{v}} +
		 2{\bb{\Omega_P}}\times {\bb{v}}\right) = - \del P - { \rho}\, \nabla \Phi .  \,\label{eq:motp}
	\end{equation}
	Here, $P$ is the pressure, $\rho$ is the density, $\Phi$ is sum of the gravitational potential due to the central mass $M_*$, $\Phi_{gr} = -GM_*/|{\bf r}|$ and the centrifugal potential, 
$\Phi_{rot}= -\Omega^2_P |{\bf r}\btimes {\hat {\bf k}}|^2/2, $ with ${\bf r}$ being the position vector measured from the central star. 
The fluid velocity is ${\bb{v}}$. (For a full list of symbols see Table \ref{symbols}.) It is expected that the evolution of the dust particle distribution can be modelled as a pressureless fluid, which has a frictional interaction with the gas as long as the dimensionless parameter $\Omega_P \tau_s\ll 1$   so that the dust is tightly coupled to 
 it \citep[see eg.][]{Garaud2004}. In the limit $\tau_s \rightarrow 0$ the system reduces to a single combined fluid in which the density may vary on account of a frozen-in dust distribution.

	We consider vortices with small length scale, such that with reference to the sound speed in the gas, relative velocities are highly subsonic. Under these conditions we expect the fluid to move incompressibly. Then we have 
	\begin{equation}
		\nabla\cdot {\bb{v}} =0 \label{eq:crppp}
	\end{equation}
	and hence 
	\begin{equation}
		\hspace{2mm}
		\frac{\partial \rho}{\partial t} + {\bb{v}}\cdot\nabla \rho = 0 .\,\label{eq:contdspp}
	\end{equation}
	The density is thus conserved for fluid elements   corresponding to a frozen-in dust distribution.
         Dissipative processes would cause this distribution to evolve slowly in time.
         However, in this paper for simplicity we shall assume any assosiated  time scale is much longer than evolutionary tine  scales of interest, such as those associated with dynamical
	instabilities. Thus we adopt equations (\ref{eq:motp}), (\ref{eq:crppp}) and (\ref{eq:contdspp}) throughout.

	\section{Steady state solutions } \label{sec:steady_state_equations}
	In a steady state, the equation of motion (\ref{eq:motp}) reduces to 
	\begin{equation}
		 {\bf v}\cdot \nabla{\bf v} + 2\bmth{\Omega_P}\times {\bf v}
		= -\frac{\nabla P} {\rho} - \nabla \Phi. \label{eq:1sst}
	\end{equation}
	In order to consider local steady state solutions within a Keplerian disc in detail, we adopt a local shearing box with origin centred on a point of interest and rotating with 
its Keplerian angular velocity \citep[see][]{Goldreich1965,Regev2008}. This specifies $\Omega_P$.
 A local Cartesian coordinate system is adopted with the $x$-axis in the radial direction, the $y$-axis in the direction of shear 
and the $z$-axis normal to the disc mid-plane. For a general vector ${\bf a}$ we adopt the equivalent representations
 ${\bf a}\equiv (a_x, a_y, a_z) \equiv (a_1, a_2, a_3).$ To within an arbitrary constant and up to order $x^2,$
 the combined centrifugal and gravitational potential $\Phi = -\Omega_P^2(3 x^2-z^2)/2$. The length scale associated with each dimension of the box can be taken to be the vertical scale height which, in the thin disc approximation, is assumed to be very much less than the local radius or distance to the central mass.

\subsection {Solutions that are independent of $z$ }\label{sec:indepz}
	We look for solutions of equations (\ref{eq:crppp})-(\ref{eq:1sst}) for which the fluid state variables are independent of $z$ and have $ v_z\, \equiv v_3 = 0$. In order to do this the $z$ dependence of $\Phi$ is ignored. In order to satisfy the condition $\nabla\cdot\, {\bf v}=0,$ 
	we adopt a stream function $\psi,$ such that ${\bf v}\, = ({\partial \psi}/{\partial y}, -{\partial \psi}/{\partial x}, 0)$. For the undisturbed background Keplerian flow, ${\bf v}\, = (0, -3\Omega_P x/2, 0)$ and thus $\psi = \psi_0 = 3\Omega_P x^2/4$. 

	Although these solutions so not depend on $z$, we remark that they may apply to horizontal planes of an isothermal disc for which hydrostatic equilibrium holds in the $z$ direction \citep[see also][]{Lesur2009}. In that case $\rho \propto \exp(-(\Omega_P^2z^2/(2c_s^2)),$ 
	where $c_s$ is the constant local isothermal sound speed and in the thin disc approximation the vertical scale height  $ H = c_s/\Omega_P \! \ll\! r$ is implicit \citep[see e.g.][]{Pringle1981}. It is readily seen that a factor $\propto exp(-\Omega_P^2z^2/(2c_s^2))$ may be applied to the two dimensional solutions for $ \rho$ and $P$ obtained from (\ref{eq:crppp})-(\ref{eq:1sst}). 
	Then when the $z$ dependence is restored to $\Phi,$ hydrostatic equilibrium will hold in the $z$ direction. Note that  this feature  depends on there being  no   $x$   dependence in the above expression for $\rho$, which in turn depends on the adoption of the quadratic potential  $\Phi = -\Omega_P^2(3 x^2-z^2)/2,$ which is valid in the thin disk limit to within a correction of order $H/r$. The characteristic velocity associated with the box being $c_s$  in the thin disc limit on dimensional grounds, the characteristic velocity associated with this correction is then expected to be  of order $c_sH/r.$ This may be assumed to be small for thin enough discs even for vortices with subsonic velocities.  We remark that the above description of local solutions in vertical hydrostatic equilibrium, with negligible vertical flows in the thin disc limit, has been found numerically  to be applicable to vortices generated by the Rossby wave instability \citep[see eg.][]{Lin2012}. Note too that in the limit of zero stopping time considered here, the dust is frozen into the fluid so that vertical settling does not occur.

\subsection{Functions specifying the vorticity and density profiles} \label{sec:specifying_profiles}
	For a steady state, equation (\ref{eq:contdspp}) becomes $ {\bb{v}}\cdot\nabla \rho = 0.$ For a two dimensional flow this implies that the density is constant on streamlines and is thus a function of $\psi$ alone. Accordingly we may write, $\rho =\rho(\psi),$ where $\rho(\psi)$ is an arbitrary function of $\psi$. This cannot be determined if $\tau_s=0 $ as it is then an invariant that must be input externally. It may be considered to be the result of evolutionary processes taking place on a long time scale 
	when the condition $\tau_s=0$ is relaxed.
	The steady state equation of motion (\ref{eq:1sst}) may be recast in the form
	\begin{equation}
		 \left( 2\bmth{\Omega_P}+ \boldsymbol{\omega}\right)\times {\bf v}
		= -\frac{ P}{\rho^2}\nabla \rho 
		- \nabla \left(\frac{ P}{{\rho}} + \Phi + \frac{1}{2}\vert \bf{v}\vert^2\right), \label{eq:1sstp1} 
	\end{equation}
	where $\omega_z={\hat{\bf k}}\cdot\nabla\times { \mathbf{v}} = - \nabla^2\psi$ is the $z$ component of 
         vorticity, $\bmth{\omega},$ as observed in the rotating frame. Expressing quantities in terms of $\psi,$ 
         equation (\ref{eq:1sstp1}) becomes 
	\begin{equation}
		\left(-\nabla^2\psi + 2\Omega_P + \frac{P}{\rho^2} \frac {d \rho}{d\psi}\right)
		\nabla \psi = -\nabla \left( \frac{P}{\rho} + \Phi + \frac{1}{2}
		{\vert \nabla \psi \vert}^2 \right) \equiv -\nabla F \label{eq:fq}
	\end{equation}
	As both sides of the above   have to be 
        proportional to $\nabla \psi,$ it follows that $F$ is a function of $\psi$ alone, or $F=F(\psi).$ 
        In the absence of diffusive processes, this arbitrary function, the derivative of which in the absence of a density gradient represents a conserved vorticity, also has to be input externally.

	Equation (\ref{eq:fq}) can thus be written as a second order partial differential equation for the stream function in the form
	\begin{equation}
		\nabla^2\psi = \frac{d F}{d\psi} + 2\Omega_P + \frac{P}{\rho^2} \frac {d \rho}{d\psi}. \label{eq:fq1}
	\end{equation}
	We remark that once $F(\psi)$ is specified, the pressure is expressed in terms of the stream function through the relation 
	\begin{equation}
		\frac{P}{\rho}= - \Phi - \frac{1}{2}{\vert \nabla \psi \vert}^2+ F(\psi). \label{Pcalc}
	\end{equation}
	Solutions of (\ref{eq:fq1}) corresponding to local vortices with central dust concentrations may be sought once the arbitrary functions $\rho$ and $F$, 
       and appropriate boundary conditions are specified. In this context we note that, after making an appropriate adjustment to $F$,
        equation (\ref{eq:fq}) is invariant to adding an arbitrary constant to $P$. For convenience, when constructing  steady state vortices numerically, 
        we shall choose this to make the pressure zero in the limit when the flow becomes a pure Keplerian background flow with no added vortex. 

\subsection{Specification of $F(\psi)$ and $\rho(\psi)$ in practice}\label{sec:sol_w_dust_concentrations}
	We begin by separating out the solution corresponding to an undisturbed Keplerian background flow for which $\psi=\psi_0$. 
	To do this we write 
	\begin{align}
		\psi&=\psi_0+\psi_1=\frac{3}{4}\Omega_Px^2+\psi_1,
	\end{align}
	where $\psi_1$ corresponds to the superposed vortex. In addition we set
	\begin{align}
		F=-\frac{\Omega_P\psi
                   }{2} +F_1(\psi),
	\end{align}
	where $F_1$ vanishes for the background flow. Equation (\ref{eq:fq1}) then yields
	\begin{equation}
		\nabla^2 \psi_1 = \frac{d F_1}{d \psi}+ \frac{P}{\rho^2}\frac{d \rho}{d\psi} =
		\mathcal{A}(\psi)+\frac{P}{\rho}\mathcal{B}(\psi). \label{eq:psi1}
	\end{equation}
	Here we denote $\mathcal{A}(\psi)$ as the Bernoulli source and $\mathcal{B}(\psi)$ as the density source, these both being regarded as sources of vorticity. 

	As in the limit $\tau_s = 0,$ these functions are invariants that have to be specified. Accordingly we specify $\mathcal{A}(\psi)$ and $\mathcal{B}(\psi)$ so as to enable a large class of steady state solutions with varying vorticity and density profiles to be considered. The Bernoulli and density sources are superposed on horizontal planes on which there is a uniform background Keplerian flow with density $\rho_0.$ They are non zero only on streamlines that circulate interior to a bounding streamline, with the location of the point where it crosses the $y$ axis being specified. The arbitrary unit of length is chosen so that this point is at $(0,1),$ the ignorable $z$ coordinate being from now on suppressed. The configurations are symmetric with respect to reflections in both the $x$ and $y$ axes. The unit of time is chosen so that $\Omega_p =1$, while the arbitrary unit of mass is then chosen so that $\rho_0=1.$ In order to perform calculations we adopt power law functions of the form
	\begin{IEEEeqnarray}{rCl}
		 \mathcal{A}(\psi)&=&A\vert\psi-\psi_{b}\vert^{\alpha} \label{eq:A}\\
		\rho(\psi)-\rho_0&=&B\vert\psi-\psi_{b}\vert^{\beta} .\label{eq:RHO}
	\end{IEEEeqnarray}
	Here $\psi_{b}$ denotes the value of $ \psi$ evaluated on the vortex core boundary which intersects $(0,1).$
 The functions $\mathcal{A}(\psi)$ and $\rho(\psi)-\rho_0$ are set to be zero on streamlines exterior to those with with $\psi = \psi_{b}$. 
The constants $A$ and $B$ are chosen to scale the total vorticity and relative mass excesses associated with the Bernoulli and mass sources respectively
 and $\alpha$ and $\beta$ are constant indices. Note that $A>0$ as considered here gives rise to an anticyclonic vortex.
 In particular when $\alpha=\beta =B=0$ we obtain the well known Kida solution \citep{Kida1981,Lesur2009}. The specification of the vorticity sources through the above procedure ensures  that they vanish on and exterior to the vortex boundary. In addition, solutions covering a wide range of vortex aspect ratios with a variety of density and vorticity profiles may be obtained by varying the constants $A$ and $B$ and the indices $\alpha$ and $\beta$.

\subsection{The Kida solution}\label{sec:kidavortex}
	\begin{figure*}
	 \begin{minipage}{150mm}
	 \centering
	 \includegraphics[width=15cm]{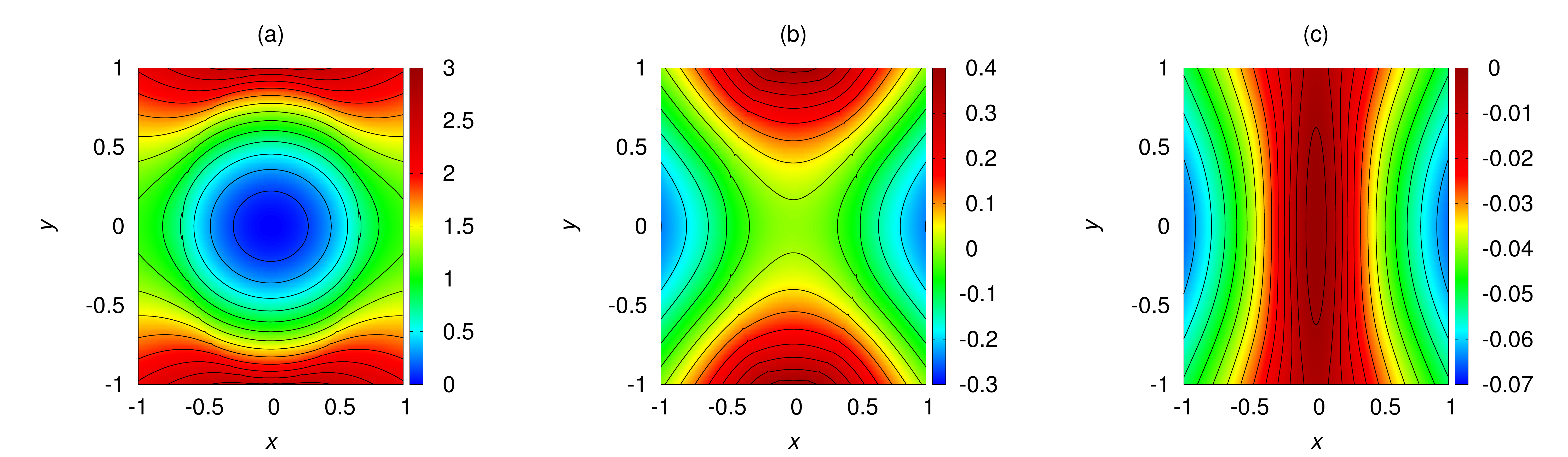}
	 \caption{The pressure distributions inside Kida vortices with a range of aspect ratios, $\chi$. In (a), the aspect ratio $\chi=3/2$ and the vortex contains a pressure minimum in its centre. In (b), $\chi=2$ and there are hyperbolic pressure contours inside the vortex and a saddle point at its centre. In (c) $\chi=5$ and there is a pressure maximum at the centre of the vortex. Theses correspond to a $\omega_m$ of $3.33, 1.125$ and $0.09$ respectively. All vortices with $\chi > 4$ have pressure maxima at their centre. Note that in this and subsequent Figures, an arbitrary constant can be added to the pressure in order to avoid negative values.}
	 \label{fig:kidapressures}
	 \end{minipage}
	\end{figure*}
	The Kida vortex provides a well-known 2D steady state analytic solution. The core is an elliptical patch with constant vorticity $\omega_t =-S + \omega_v$. Here $-S$ is the vorticity associated with the background flow as seen in the rotating frame, with $S= 3\Omega_P/2 $ in the Keplerian case, while $\omega_v$ is the vorticity imposed on the background. Outside the core, the vorticity is that of the background flow. This corresponds to a Bernoulli source inside the core given by $\mathcal{A}(\psi)=-\omega_v$ in the above notation.
	 
	Kida vortices are characterised by their aspect ratio $\chi=a/b \ge 1$, where $a$ and $b$ are the semi-major and semi-minor axes of the elliptical streamlines within the core. One can find steady solutions when the semi-major axis of the vortex is aligned with the background shear and we have
	\begin{equation}
		\omega_v=-\frac{S( \chi+1)}{\chi(\chi-1)}.
	\end{equation}
	The stream function in the vortex core is given by 
	\begin{equation}
		\psi = \psi_0 +\psi_1 = \frac{S }{2(\chi-1)}\left(\chi x^2+ \frac{y^2}{\chi}\right), \label{Kidasol}
	\end{equation}
	This is found by looking for a solution with elliptical streamlines  which is  connected  to an exterior solution of Laplace's equation  under the condition of continuity of  $\psi$ and $\nabla \psi$ on the vortex  boundary \citep[see][ and references therein for more details ]{Lesur2009}. From equation (\ref{Pcalc}), for a fixed value of $z$, the pressure is then given to within an arbitrary constant by
	\begin{equation}
	\frac{P}{\rho}= \frac{3\Omega_P^2x^2}{2} - \frac{S^2}{2(\chi-1)^2}\left(x^2\chi^2 +\frac{y^2}{\chi^2}\right)
	+ \frac{S\psi( \chi^2 +1)}{\chi(\chi-1)}-2\Omega_P\psi, \label{Pcalc1}
	\end{equation}
	where the last two terms constitute  $F(\psi)$. For a Keplerian disk with uniform background $S=3\Omega_P/2$. In that case there are a range of pressure profiles associated with different aspect ratios as can be seen in Figure~\ref{fig:kidapressures}. This is particularly relevant when considering dusty gases as particles tend to drift in the direction of the pressure gradient towards pressure maxima.

\subsection{An analytic polytropic model with variable density}\label{polyeq}
	We remark that it is possible to consider different values of $S$ (i.e. a non-Keplerian background flow) while retaining the potential $\Phi$ that is appropriate for a Keplerian disc. This requires the  pressure gradient to be non zero in the background flow and as a consequence enables  us to consider situations where the vortex is centred on a background where there is a pressure extremum. We comment that this is of interest  as dust is expected to  accumulates at the centre of a ring where there is a pressure maximum \citep{Whipple1972} and in addition,  the Rossby wave instability can result in  vortices forming  at  such locations \citep[see e.g.][]{Meheut2010a,Meheut2012a}.
 
	For example if we set
	\begin{equation}
		S=\sqrt{\frac{3(\chi-1)}{(\chi+1)}}\Omega_P, \label{polysh}
	\end{equation}
	then we have to within a constant in the vortex core that
	\begin{equation}
		\frac{P}{\rho}= -\frac {S\psi}{\chi(\chi-1)}+F(\psi).\label{polyp}
	\end{equation}
	This makes $P/\rho$ a function of $\psi$ alone which will be a linear function of $\psi$ provided that $F(\psi)$ is. This turns out to be useful for constructing models with non-uniform density. However, when such a solution is matched to an exterior solution \citep[e.g.][]{Lesur2009} the background flow will correspond to one with $v_y= -Sx$. From (\ref{polysh}) this corresponds to the Keplerian case strictly only when $\chi=7$. For other values of $S,$ consideration of the exterior Kida solution implies that there is an implied background pressure structure which corresponds to a background pressure maximum at the coorbital radius for $\chi >7$ and a background pressure minimum there for $\chi < 7$.
	
Developing the above discussion further, we obtain an analytic model with variable density. This will have a stream function of the form (\ref{Kidasol}) with $S$ given by (\ref{polysh}) inside the vortex core where the vorticity source will be uniform. To obtain this solution we set $\rho=(1-b\,(\psi-\psi_b)/\psi_b)^{n_1}$ inside the vortex, where $\psi_b$ is the stream function on the core boundary and the background density is taken to be unity. The quantities $b$ and $n_1$ are constants determining the profile and magnitude of the density excess above the background. At the vortex centre $\rho=(1+b)^{n_1}$ while at the boundary $\rho=1$, the background value. The pressure is assumed to take the form $P= \beta_P\,\psi_b\,\rho^{1+1/n_1}/((n_1+1)\,b)$, where $\beta_P$ is a constant determined such that the equilibrium conditions apply. These conditions are obtained from (\ref{eq:fq1}) and (\ref{polyp}) which require that $F$ is a linear function of $\psi$. They give
	\begin{IEEEeqnarray}{rCl}
	\frac{S(\chi^2+1)}{\chi(\chi-1)}&=&\frac{dF(\psi)}{d\psi}+2\Omega_P-\frac{n_1\beta_P}{n_1+1}\hspace{3mm} {\rm and} \nonumber\\
	\frac{\beta_P\,\psi_b}{(n_1+1)\,b}\left(1-\frac{b(\psi-\psi_b)}{\psi_b}\right)&=&-\frac{S\psi}{\chi(\chi-1)}+F(\psi).
	\end{IEEEeqnarray}
	Together they imply that
	\begin{equation}
	\beta_P= \Omega_p\left(2-\chi\sqrt{\frac{3}{(\chi^2-1)}}\,\right),
	\end{equation}
	which determines $\beta_P.$ The parameter $n_1 > 0$ can be specified arbitrarily. Then $b$ can then be chosen to 
	scale the density excess above the background in the centre of the vortex provided $\chi > 2.$ In this paper we have limited consideration
	to the case $n_1=1.$

\section{The stability of general incompressible vortical flows allowing for density gradients }\label{stability}
	We now consider the stability of steady state flows of the type introduced above. We find it useful to consider both the Eulerian and Lagrangian formulation of the linear stability problem as they are found to be convenient for different purposes. Following the Lagrangian approach developed by \citet{Lynden-Bell1967}, we introduce the Lagrangian variation $\Delta$ such that the change to a state variable $Q$ as seen following a fluid element is $ \Delta Q$. The Lagrangian displacement is given by $\Delta {\bf r} = \bmth{\xi}$ and we have 
	\begin{equation}
		\Delta {\bf v} = \frac{D \bmth{\xi} }{ D t},
	\end{equation}
	where $D/Dt$ denotes the convective derivative for the unperturbed flow. Thus
	\begin{equation}
		\frac{D}{ D t} \equiv \frac{\partial}{ \partial t}+ {\bf v}\cdot \nabla. \nonumber 
	\end{equation}
	The Eulerian variation, $Q'$, the change in $Q$ as seen in a fixed coordinate system, is given by $Q'= \Delta Q - \bmth{\xi}\cdot \nabla Q$. Taking the Lagrangian variation of the
	equation of motion (\ref{eq:motp}), we obtain
	\begin{equation}
		\frac{D^2\bmth{\xi}}{ D t^2} + 2\bmth{\Omega_P}\times \frac{D \bmth{\xi}}{ D t}
		= \Delta {\bf F}, \label{1sem} 
	\end{equation}
	with
	\begin{equation}
		\Delta {\bf F}= -\frac{\nabla P'}{ \rho} +\frac{\rho'}{\rho^2}\nabla P -\bmth{\xi}\cdot\nabla\left(\frac{\nabla P}{ \rho} + \nabla \Phi\right). \label{Feq}
	\end{equation}
	The Lagrangian variation in the density is zero, thus
	\begin{equation}
		\Delta \rho = \rho'+\bmth{\xi}\cdot\nabla\rho =0. \label{Drho}
	\end{equation}
	Equations (\ref{1sem}) and (\ref{Drho}) together with the incompressibility condition $\nabla\cdot \bmth{\xi}=0$, 
	lead to system of equations for the horizontal components of $\bmth{\xi}$ that is fourth order in time (see below). While the above Lagrangian formulation is convenient for some aspects such as the  analytic discussion of  saddle point instability  in Section \ref{saddlepoint}, the  Eulerian formulation presented  below is found to be  more convenient in other contexts.

\subsection{Eulerian form}
	The corresponding equations in terms of the Eulerian variations are
	\begin{IEEEeqnarray}{rCl}
		\frac{D {\bf v}'}{Dt}+2\bmth{ \Omega_P}\times {\bf v}'+{\bf v}'\cdot \nabla {\bf v}&=&-\frac{1}{\rho}\nabla P' +\frac{\rho'}{\rho^2}\nabla P\hspace{3mm} {\rm and} \nonumber \\
		\frac{D \rho'}{Dt}&=&-{\bf v}'\cdot\nabla\rho, \label{Eulerianp}
	\end{IEEEeqnarray}
	which is a system that is third order in time. It is lower order than the Lagrangian system on account of trivial solutions corresponding to a relabelling of fluid elements being present in the latter case \citep[see][]{Friedman1978}. Equations expressed in terms of Eulerian variations were found  to be simpler to use when analysing  vertical stability in Section~\ref{vertical_stab} and  for the same reason were  solved numerically when considering vortex stability in Section~\ref{Stabcalc}.

\subsection {Local Analysis}\label{local}
	We consider perturbations that are localized on streamlines which have short wavelengths in the directions perpendicular to the unperturbed velocity, but can have a long wavelength in the direction of the unperturbed velocity. The latter is a natural outcome of shearing motions. To do this we begin by assuming that any perturbation quantity takes the form 
	\begin{equation}
		\Delta Q = \Delta Q_0 \exp\,( i\lambda S_{\!A}). \label{locco}
	\end{equation}
	Here we adopt a WKBJ ansatz with the phase function $S_{\!A}$ left arbitrary for the time being and the constant $\lambda$ taken to be a large parameter. 
	For a discussion of the approach followed here in a variety of contexts, see \citet{Lifschitz1991}, \citet{Sipp2000} and \citet{Papaloizou2005}.
	The effective wavenumber 
	\begin{equation}
	{\bf k} = \lambda \nabla S_{\!A} \label{waveno}
	\end{equation}
	then has a large magnitude.
	The amplitude factor $\Delta Q_0$ is the WKBJ envelope.
	Due to the rapid variation of the complex phase $\lambda S_{\!A},$ one can perform a WKBJ analysis, such
	that the state variables $\Delta Q_0$ are expanded in inverse powers of $\lambda.$ The lowest order term in $\bmth{\xi}$ is constant, while the lowest order term in $P'$ is $\propto \lambda^{-1}.$ To lowest order (\ref{1sem}) gives
	\begin{equation}
		\frac {D S_{\!\!A}}{Dt} =0.\label{ikonal}
	\end{equation}
	When working to the next order, only the variation of the rapidly varying phase $S_{\!A}$ needs to be considered when taking spatial  derivatives, apart from when considering expressions involving the operator $D/Dt\equiv \partial/\partial t+{\bf v}\cdot \nabla$ as this annihilates $S_{\!A}$. Accordingly the contribution ${\bf v}\cdot \nabla (\Delta Q_0)$ must be retained. Noting the above, we can substitute perturbations of the form (\ref{locco}) into the governing equations and remove the factor $\exp\,( i\lambda S_{\!A}),$ thus obtaining equations for the lowest order contribution to the quantities $\Delta Q_0$ alone. For ease of notation we drop the subscript $0$ from now on.

	Following this procedure (\ref{1sem}) gives
	\begin{equation}
		\frac{D^2\bmth{\xi}}{ D t^2}+ 2\bmth{\Omega_P}\times \frac{D \bmth{\xi}}{ D t}
		= -\frac{\ii\!  {\bf k} P'}{ \rho} - \frac{\bmth{\xi}\cdot\nabla\rho}{\rho^2}\nabla P +\bmth{\xi}\cdot\nabla\left({\bf v}\cdot \nabla {\bf v} + 2\bmth{\Omega_P}\times {\bf v} \right), \label{2sem} 
	\end{equation}
	where we recall that the lowest order term for $P'\propto \lambda^{-1}$. In addition to this, the incompressibility condition gives
	\begin{equation}
		{\bf k}\cdot\bmth{\xi}=0. \label{2scm}
	\end{equation} 
	Using (\ref{waveno}) and (\ref{ikonal}) we can also find an equation for the evolution of ${\bf k}$ in the form
	\begin{equation}
		\frac{D \bmth{k}}{ D t}=-k_j\nabla v_j. \label{kdot}
	\end{equation}
	Equations (\ref{2sem}), (\ref{2scm}) and (\ref{kdot}) give a complete system for the evolution of $\bmth{\xi}$ and ${\bf k}$, after the elimination of $P'$, as an initial value problem. Because the evolution consists of advection of data along streamlines, it is possible to consider disturbances localized on individual streamlines \citep{Papaloizou2005}. Localization amplitudes are unaffected by the evolution considered. In general one could start with an arbitrary initial $S_{\!A}$ and then ${\bf k}$ would depend on time.

\subsubsection{Solutions for a time-independent  wavenumber}\label{timeind}
	A relatively simple class of solutions for ${\bf k}$ can be obtained by setting $S_{\!A}$ to be independent of time and a function of quantities conserved on unperturbed streamlines so that ${\bf v}\cdot \nabla S_{\!A}=0$ \citep{Papaloizou2005}. Then from the Eulerian viewpoint,
        $\bmth{k}$ is fixed for all time and we only have to solve for $\bmth{\xi}.$ For the simple case of a two dimensional vortex with initial state-independent of $z$, we may take
	\begin{equation}
		S_{\!A}= g(\psi)+k_z z/\lambda. \label{steadyk} 
	\end{equation}
	Here $\psi$ is the unperturbed stream function, $g$ is an arbitrary function
	and $k_z$ is the constant vertical wavenumber. We assume for now that $k_z\ne 0$, appropriate to the physically realistic case where perturbations are localized in $z$.

	We further remark that although the above form of ${\bf k}$ is not the most general solution of (\ref{kdot}), apart from when the velocity is linear in the coordinates as for the Kida vortex,
        other solutions are such that the magnitude of the wavenumber ultimately increases linearly with time.
        In that situation we expect that although there may be temporary amplification, the system may  not ultimately show  growth of linear perturbations exponentially with time.. 
       This situation is already well known in the context of the shearing box \citep[see][]{Goldreich1965}.

	For now we continue the discussion adopting the time-independent  form of ${\bf k}$ derived from $S_{\!A}$
	given by (\ref{steadyk}) and return to the discussion of more general ${\bf k}$
	and the special nature of the Kida vortex in Section \ref{Kidas} below.

	We may use the vertical component of (\ref{2sem}) together with (\ref{2scm}) to eliminate $P'$ and $\xi_z$
	and thus obtain a pair of equations for $(\xi_x, \xi_y)\equiv (\xi_1, \xi_2)$. These can be written in the form
	\begin{equation}
		\left(\delta_{ij}+ \frac{k_ik_j}{k_z^2}\right) \frac{D^2\xi_j}{ D t^2} +\left( 2\epsilon_{i3j}\Omega_P + \frac{2k_i}{k_z^2}\frac{D k_j}{ D t}\right)\frac{D \xi_j}{ D t}+\frac{k_i\xi_j}{k_z^2}\frac{D^2 k_j}{ D t^2}
		=H_i \label{3sem}
	\end{equation}
	where
	\begin{equation}
		{\bf H} = - \frac{\bmth{\xi}\cdot\nabla\rho}{ \rho^2}\nabla P +\bmth{\xi}\cdot\nabla\left({\bf v}\cdot \nabla {\bf v}+ 2\bmth{\Omega_P}\times {\bf v} \right).
	\end{equation}
        We remark that the neglect of vertical  stratification in this calculation can be generally justified if it is assumed that  $k_z^2/(k_x^2+k_y^2)$ is large, otherwise
        the modes can be assumed to be localized in the vicinity of the midplane where the vertical startification is least.

	Note that as only horizontal components are considered, the summation is for $j=1,2$ only. In addition, we readily find from (\ref{kdot}) that
	\begin{equation}
		\frac{D^2 k_j}{D t^2}= \frac{\partial }{\partial x_q} \left[\,k_{\mu}\left( v_\mu \frac{\partial v_q}{\partial x_j}  -v_q \frac{\partial v_\mu}{\partial x_j}\right)\, \right]. \label{kddot} 
	\end{equation}
	We remark that although there is an arbitrary function $g$ in the definition of ${\bf k}$ used in this section, because the derivatives in (\ref{3sem}) correspond to advecting around streamlines and $dg/d\psi$ is constant on streamlines, the latter quantity  effectively behaves as a multiplicative constant merely scaling the magnitude of the wavenumber.

\subsubsection{Eulerian form}
	The equivalent of (\ref{3sem}) written in terms of the Eulerian variations $(v_x',v_y')\equiv (v_1',v_2')$ obtained from (\ref{Eulerianp}) is
	\begin{equation}
		\left(\delta_{ij}+ \frac{k_ik_j}{k_z^2}\right) \frac{Dv'_j}{ D t}+ 2\epsilon_{i3j}\Omega_P v'_j+v'_j\frac{\partial v_i}{\partial x_j}+\frac{k_i v'_j}{k_z^2}\frac{D k_j}{ D t}
		=\frac{\rho'}{\rho^2}\frac{\partial P}{\partial x_i}.  \label{3sEem} 
	\end{equation}
	In addition, the Eulerian density variation satisfies
	\begin{equation}
		\frac{D\rho'}{D t}= -{\bf v}'\cdot\nabla\rho. \label{3sEce} 
	\end{equation}

\subsubsection{Generic instability}
	The analyses in the previous sections reduce the stability problem to solving an initial value problem of integrating a set of simultaneous ordinary differential equations around streamlines. The independent variable measures the location on a streamline.

	As the unperturbed motion on a streamline is periodic, these equations have periodic coefficients through their dependence on ${\bf k}$ and the gradient of $\rho$ etc. Thus Floquet theory may be applied \citep[eg.][]{Whittaker1996}. According to this, if some internal mode with a natural oscillation frequency dependent on ${\bf k}$ is described by (\ref{3sem}), unstable bands of exponential growth are expected as ${\bf k}$ is varied to allow resonances of this frequency with the frequency of motion around the streamline.
	To see how this can come about we shall specialise to the case when $|k_z| \gg \sqrt{k_x^2+k_y^2}$ which corresponds to the so called horizontal instability, as in this limit the motion occurs in uncoupled horizontal planes.

\subsection{Horizontal instability}
	Taking the limit $k_z \rightarrow \infty$, the horizontal components of (\ref{3sem}) yield a pair of equations for the horizontal components of displacement, 
        while the $z$ component and the condition $\nabla\cdot \bmth{\xi}=0$ yield $\xi_z \rightarrow 0$. We thus have
	\begin{equation}
		\frac{D^2\bmth{\xi}}{ D t^2}+ 2\bmth{\Omega_P}\times \frac{D \bmth{\xi}}{ D t}
		= - \frac{\bmth{\xi}\cdot\nabla\rho}{\rho^2}\nabla P +\bmth{\xi}\cdot\nabla\left({\bf v}\cdot \nabla {\bf v}+ 2\bmth{\Omega_P}\times {\bf v} \right), \label{2sem1} 
	\end{equation}
	where we now have $\bmth{\xi}=(\xi_x,\xi_y).$ Using equation (\ref{eq:1sst}), this can be written in the equivalent form
	\begin{equation}
		\frac{D^2\bmth{\xi}}{ D t^2} + 2\bmth{\Omega_P}\times \frac{D \bmth{\xi}}{ D t} =
		-\left( \frac{\bmth{\xi}}{\rho}\cdot\nabla\right)\nabla P -\left( {\bmth{\xi}\cdot\nabla}\right)\nabla \Phi \label{2sem2} 
	\end{equation}
	We remark that (\ref{2sem2}) becomes an equation with constant coefficients for $\bmth{\xi}$ when $\rho $ is constant and $P$ and $\Phi$ 
        are quadratic in $x$ and $y$. As this is always the case arbitrarily close to the centre of any regular vortex where there is a stagnation point, there are generic consequences.

\subsection{The saddle point instability close to the vortex centre}\label{saddlepoint}
	In the horizontal limit we solve (\ref{2sem1}) by setting $\bmth{\xi}=\bmath{\xi_0}\exp{\ii\! \sigma t}$, where $\bmth{\xi_0}$ is a constant vector, and finding an algebraic equation for $\sigma$. In doing this we find it convenient to define the symmetric matrix ${\sf E}$ by writing the right hand side of (\ref{2sem2}) as $({\sf E}(\xi_x,\xi_y)^T)^T$. The equation for $\sigma$ is readily found to be given by
	\begin{equation}
		\sigma^4 -\sigma^2(4\Omega_P^2- \Tr\,({\sf E}))+\det({\sf E}) =0, \label{disprel}
	\end{equation}
	with $\Tr$ and $\det$ denoting the trace and determinant respectively.

\begin{figure*}
		\begin{minipage}{177mm}
			\begin{center}
				\subfigure[$\left\{\alpha,\beta,\rho_m,\omega_m\right\}=\left\{0,0,0,0.09\right\}$, the Kida vortex where $\chi=5$]
				{
					\hspace{-10mm}\includegraphics[width=1.1\textwidth]{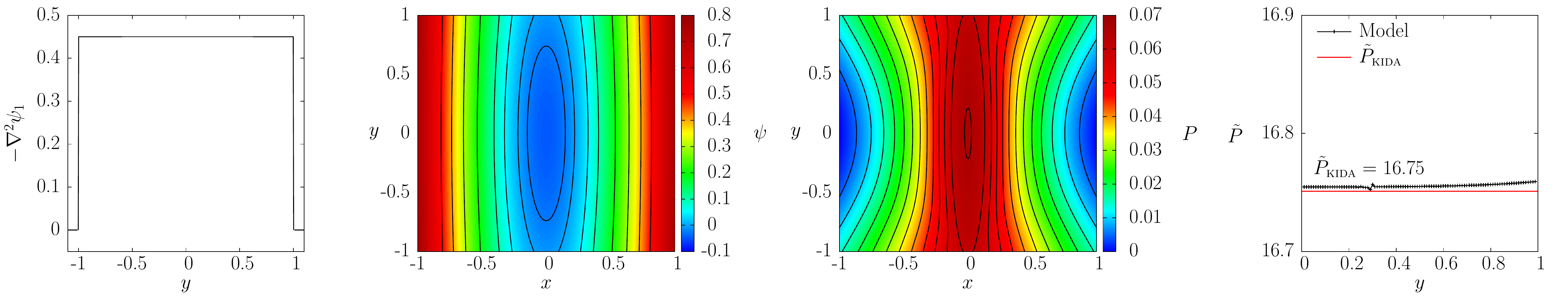}
				}
				\subfigure[$\left\{1.0,0,0,0.09\right\}$, with $\chi=4.0$.]
				{
					\hspace{-10mm}\includegraphics[width=1.1\textwidth]{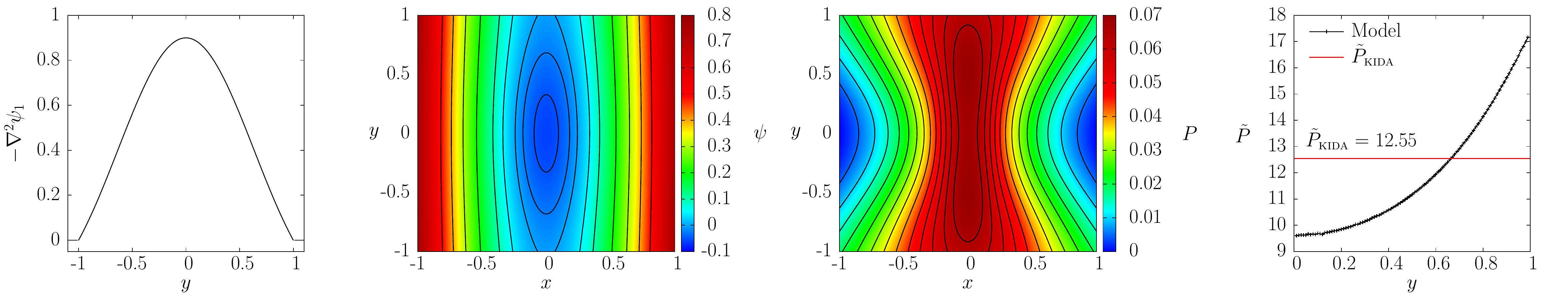}
				}
				\subfigure[$\left\{2.0,0,0,0.09\right\}$, with $\chi=3.6$.]
				{
					\hspace{-10mm}\includegraphics[width=1.1\textwidth]{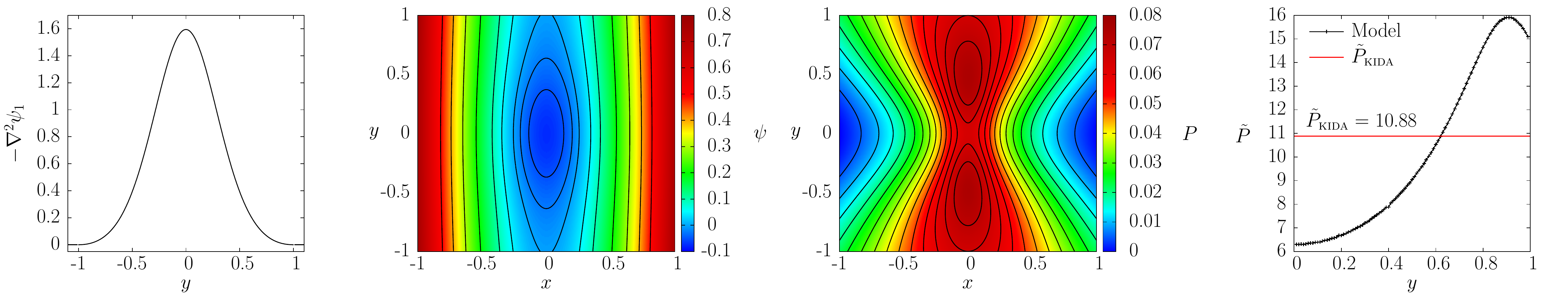}
				}
				\caption{Vortices with $\omega_m=0.09$ and density equal to the background value. From left to right we have: (i) the vorticity distribution, (ii) the $\psi$ distribution, (iii) the pressure distribution and (iv) a plot of the period $P$ around a streamline against the value of the positive $y$-coordinate where it intersects the $y$ axis. Increasing $\alpha$ from $0$ to $4$ results in a vortex that is stronger (i.e. sheared less by the background flow) and accordingly has a smaller aspect ratio. Note that all these vortices, except the one illustrated in the bottom panels, have a pressure maximum at the vortex centre. The pressure distribution in the latter vortex has a saddle point. There is also significant shear, as indicated by the variation in the period $P$ to circulate around a streamline.}
				\label{equilnodensity} 
			\end{center}
			\end{minipage}
	\end{figure*}

	A sufficient condition for instability, or at least one complex root for $\sigma$, is that $\det{\sf E}<0$. Note that in the limit approaching the vortex centre, the elements of ${\sf E}$ are given by 
	\begin{equation}
		E_{i,j}=- \frac{1}{\rho}\frac{\partial^2}{\partial x_ix_j}\left(P-\frac{3}{2}\rho\,\Omega_Px^2\right)
	\end{equation}
	evaluated in the limit $|{\bf r}| \rightarrow 0$. This condition is equivalent to $P-3\rho\,\Omega_Px^2/2$ having a saddle point at the centre.
 This will occur when $P$ has a saddle point that appears as a maximum along the $x$-coordinate line and a minimum along the $y$-coordinate line, as  can be seen to occur directly from the   analytic solution
  for  Kida vortices with $3/2 < \chi <4,$
   and for the vortex illustrated in the bottom panels of Figure~\ref{equilnodensity}. Saddle points of this type are generically associated with instability in all cases, independently of density or vorticity profile.

	\begin{figure*}
		\begin{minipage}{177mm}
			\begin{center}
				\subfigure[$\left(\alpha,\beta,\rho_m\right)= \left\{0,1.0,0.1\right\}$, with $\chi=4.93$.]
				{
					\hspace{-10mm}\includegraphics[width=1.1\textwidth]{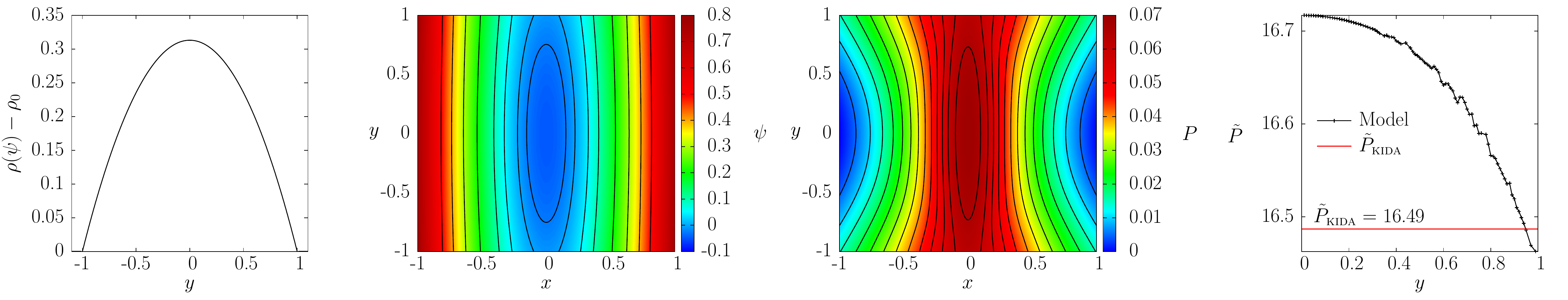}
				}
				\subfigure[$\left\{0,1.0,0.3\right\}$, with $\chi=4.84$.]
				{
					\hspace{-10mm}\includegraphics[width=1.1\textwidth]{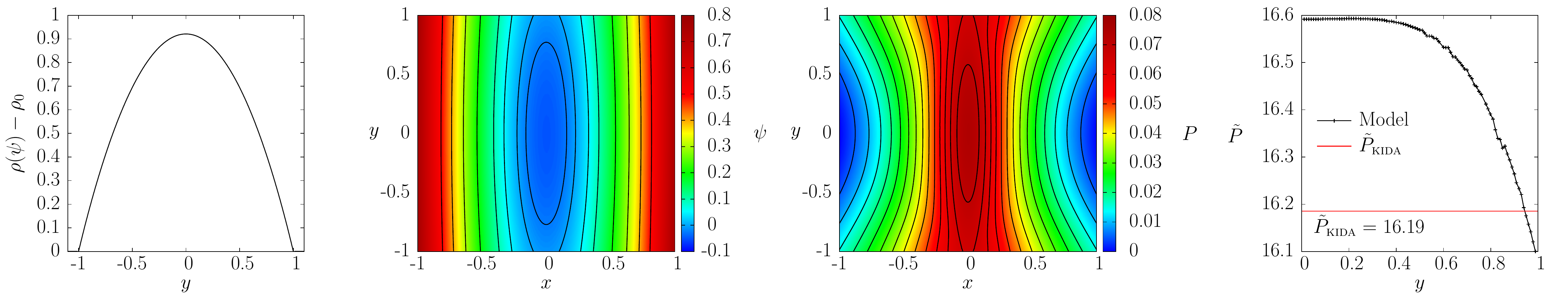}
				}
				\subfigure[$\left\{0,1.0,0.5\right\}$, with $\chi=4.80$.]
				{
					\hspace{-10mm}\includegraphics[width=1.1\textwidth]{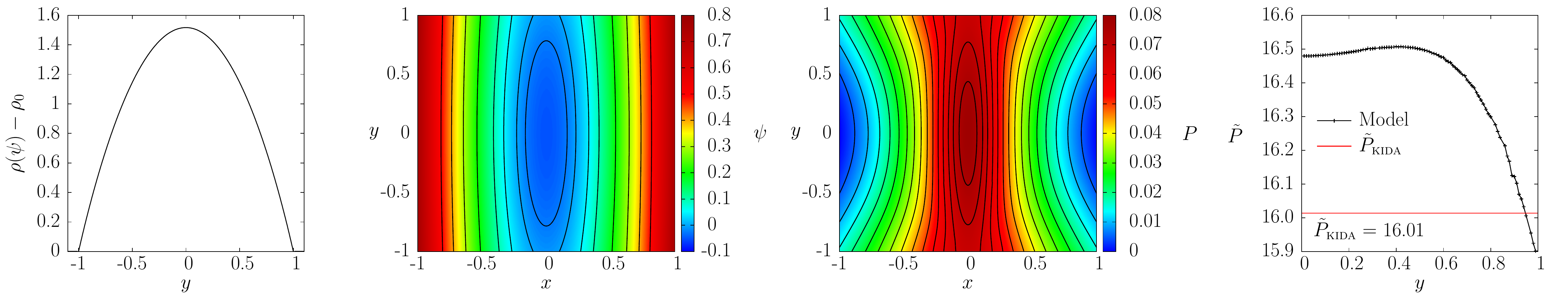}
				}
				\caption{Vortices with varying density excess for which $\omega_m=0.09$, $\alpha=0$ and $\beta=1$. From left to right we have: (i) the density distribution, (ii) the $\psi$ distribution, (iii) the pressure distribution and (iv) a plot of the period $P$ around a streamline against the value of the positive $y$-coordinate where it intersects the $y$ axis.}
				\label{equilKidadensity}
			\end{center}
		\end{minipage}
	\end{figure*}

	\begin{figure*}
		\begin{minipage}{177mm}
			\begin{center}
				\subfigure[$\left( \alpha,\beta, \rho_m,\omega_m\right) = \left\{0.5,1,0.1,0.09\right\}$, with $\chi=4.31$]
				{
					\hspace{-10mm}\includegraphics[width=1.1\textwidth]{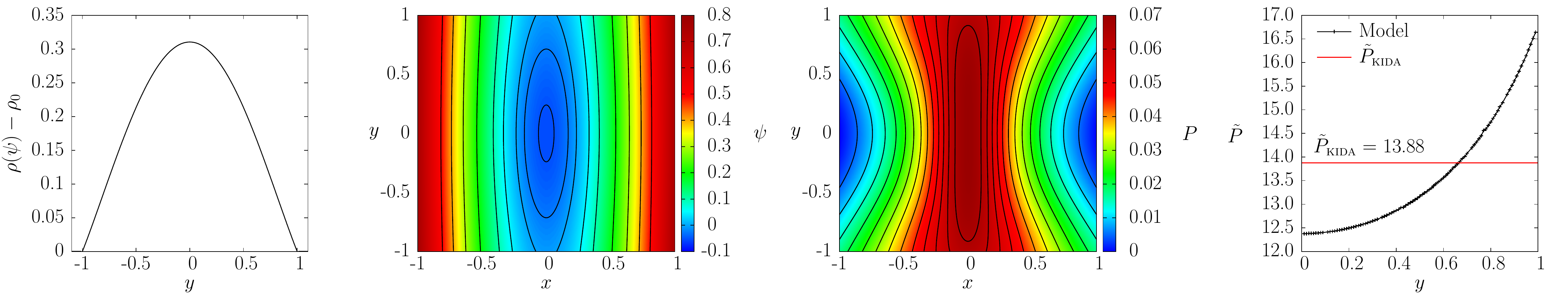}
				}
				\subfigure[$\left\{0.5,1,0.3,0.09\right\}$, with $\chi=4.23$]
				{
					\hspace{-10mm}\includegraphics[width=1.1\textwidth]{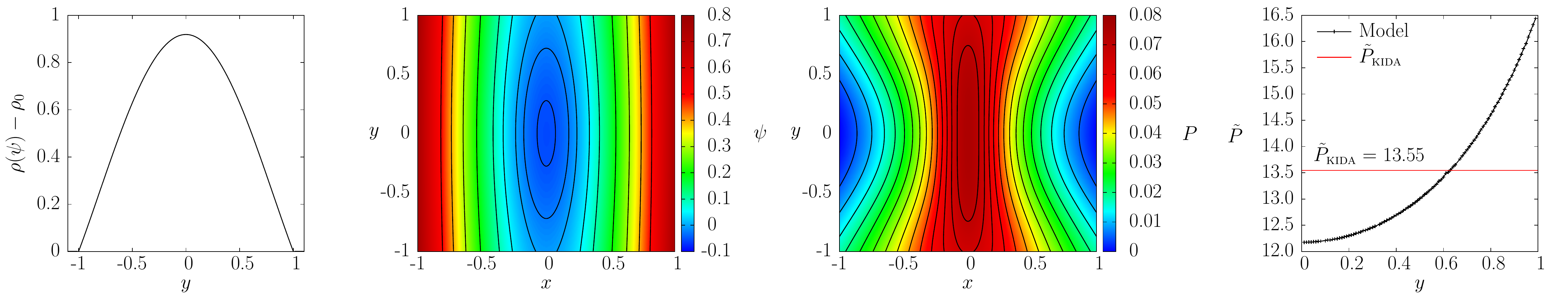}
				}
				\caption{Vortices with a non-uniform Bernoulli vorticity source , with $\omega_m=0.09$ and non-zero density enhancement parameter $\rho_m$. From left to right we have: (i) the density distribution, (ii) the $\psi$ distribution, (iii) the pressure distribution and (iv) a plot of the period $P$ around a streamline against the value of the positive $y$-coordinate where it intersects the $y$ axis. The vorticity profiles in these vortices are non uniform, resulting in significant variation of the period for circulating around internal streamlines and hence significant internal shear.}
				\label{equilnonKidadensity}
			\end{center}
		\end{minipage}
	\end{figure*}

\subsection{Parametric instability away from the vortex centre}\label{Parc}
	We recall that (\ref{2sem2}) applies in the limit $k_z\rightarrow \infty$ and that it becomes an equation with constant coefficients when $\rho$ is constant and $P$ and $\Phi$ are quadratic in the coordinates. This is always the case for any streamline in the core of a Kida vortex so moving away from the centre has no effect. However, in more general cases $P$, $\rho$ and hence ${\sf E}$ will be represented by a power series in $x^2$ and $y^2.$ Therefore for a fluid element, ${\sf E}$ will be periodic with period one half of that associated with circulating around the streamline. Thus equation (\ref{2sem2}) will have coefficients that are periodic in time and parametric instability becomes possible.

	Close to the vortex centre the time-dependence can be treated as a perturbation and parametric instability derived analytically, albeit in terms of unknown coefficients. This is done in detail in \citet{Papaloizou2005}, where a procedure is followed that can be applied directly to the problem considered here, so for the sake of brevity we refer the reader to Appendix B of that paper. 

	We note that parametric instability is first expected to occur when the epicyclic oscillation period is equal to the period to circulate around the streamline. Higher order bands are expected to be generated
	 when the ratio of epicyclic oscillation period to circulation period is $1/2,\,1/3,\dotsc$ For a vortex with a core like the Kida vortex, these resonances occur when $\chi = 4.65,\,5.89$ and $7.32$ respectively \citep{Lesur2009}.


\subsection{The stability analysis of a Kida vortex core and more general forms for the wavenumber ${\bf k}$}\label{Kidas}
	The form of wavenumber described in Section \ref{timeind} differs from that adopted in the stability analysis of the Kida vortex core given by \citet{Lesur2009}. However, because the governing equations (\ref{3sEem}), a with $\rho'=0$, are independent of the choice of the origin of time, they turn out to be equivalent. Results are illustrated in section \ref{Stabcalc}.

	Equations (\ref{2sem}) - (\ref{kdot}) of Section \ref{local} apply in this case  with  the phase function being given by $\lambda S_{\!A}={\bf{k}}(t)\cdot \bf{r}$ where the
	 wave vector ${\bf k} $ is given by
	\begin{equation}
		{\bf{k}}(t) = k_0 \Big(\chi \sin(\theta) \sin\big({\overline \phi}(t)\big), \sin(\theta) \cos\big({\overline \phi}(t)\big),\cos(\theta) \Big), \label{solution_k}
	\end{equation}
	where $k_0$ and $\theta $ are constants and ${\overline \phi}(t)=3\Omega_P/(2(\chi-1))(t-{\overline t}),$ ${\overline {t} }$ being constant. It is important to note that (\ref{solution_k}) only works when the velocity components are linear functions of the coordinates. Note that in spite of this restriction \citet{Chang2010} used it in the problem of the stability of a vortex with a density gradient, for which this condition would not be expected to be self-consistently satisfied. Thus an assessment of the situation that occurs when the velocity in the vortex is not a linear function of the coordinates should be carried out. 
 
\subsection{General form of $S_{\!A}$ for an arbitrary two dimensional incompressible vortical flow}
	The general form of $S_{\!A}$ is obtained from the solution of equation (\ref{ikonal}) in the form
	\begin{equation}
		\frac {DS_{\!A}}{Dt} =0.
	\end{equation}
	For the case when the background flow is independent of $z,$ 
    we can write $S_A=S_{\!\perp}(x,y)+k_z z/\lambda$, where $k_z$ is the component of ${\bf k}$ in the $z$ direction and $S_{\!\perp}$ satisfies
	\begin{equation}
		\frac {\partial S_{\!\perp}}{\partial t}+ \frac {\partial \psi}{\partial y}\frac {\partial S_{\!\perp}}{\partial x} - \frac {\partial \psi}{\partial x}\frac {\partial S_{\!\perp}}{\partial y} = 0.\label{ikonal1}
	\end{equation}
	The general solution of (\ref{ikonal1}) requires that $S_{\!\perp}$ be a function only of quantities that are invariant of the particle trajectories obtained by solving
	\begin{IEEEeqnarray}{rCl}
		\frac{dx}{dt} &= & \frac{\partial \psi}{\partial y} \nonumber \\
		\frac{dy}{dt} &= & - \frac{\partial \psi}{\partial x}.\label{Hamil}
	\end{IEEEeqnarray}
	The solutions for $x$ and $y$ define orbits or streamlines that are periodic in time and on which $\psi$ is constant. The period is $2\pi/\omega,$ where $\omega$ in general will be a function of $\psi$. Quantities such as $x$ and $y$ can be expressed as a Fourier series in the form
	\begin{IEEEeqnarray}{rCl}
		x &=& \sum^{\infty}_{n=-\infty} x_n(\psi)\exp\,(\ii\! n\phi), \label{Fx}\\
		y &=& \sum^{\infty}_{n=-\infty} y_n(\psi)\exp\,(\ii\! n\phi), \label{Fy}
	\end{IEEEeqnarray}
	where $\phi = \omega (t-t_0)$ and $t_0$ is a constant on an orbit that can be taken to be the time at which $x$ passes through its maximum value.

	The general solution of equation (\ref{ikonal1}), which states that $S_{\!\perp}$ is a constant on an orbit defining a streamline, is that $S_{\!\perp}$ is an arbitrary function of $\psi$ and $t_0.$ As the orbits are periodic, this function should be periodic in $t_0$ with period $2\pi/\omega$. Accordingly $S_{\!\perp}$ can also be written as a Fourier series in the form
	\begin{equation}
		S_{\!\perp} =\sum^{\infty}_{n=-\infty} C_n(\psi)\exp\,(-\ii\! n\omega t_0)= \sum^{\infty}_{n=-\infty} C_n(\psi)\exp\,(\ii\! n\,(\phi-\omega t)). \label{sumS}
	\end{equation}
	We may now find ${\bf k}=\lambda\nabla S_{\!A}$ leading to
	\begin{IEEEeqnarray}{rCl}
		k_x &=& \lambda\left(\frac{\partial \psi}{\partial x} \frac{\partial S_{\!\perp}}{\partial \psi}+ \omega \frac{\partial y}{\partial \psi} \frac{\partial S_{\!\perp}}{\partial \phi}\right) \nonumber \\
		k_y &= & \lambda \left(\frac{\partial \psi }{\partial y} \frac{\partial S_{\!\perp}}{\partial \psi} -\omega \frac{\partial x}{\partial \psi} \frac{\partial S_{\!\perp}}{\partial \phi}\right). \label{Hamil1}
	\end{IEEEeqnarray}
	In obtaining the above we note that quantities are either expressed as functions of $(x,y) \equiv \bf {r}$ or $(\phi,\psi)$ as independent variables. Transforming between these representation is facilitated by noting that ${\bf v}=\omega\partial {\bf r}/\partial \phi$ and that the Jacobian $\partial(\phi,\psi)/\partial(x,y)$ is equal to $\omega$.
	We remark that when only terms with $n=0$ are present in the sum (\ref{sumS}), $\partial S_{\!\perp} /\partial\phi=0$ and we recover the time-independent wavenumber from the Eulerian point of view, as used in Section (\ref{timeind}). 

\subsection{Wavenumber increasing with time}\label{kpropt}
	On the other hand if terms with $n\ne 0$ occur, and $d\omega/d\psi \ne 0,$ the wavenumber is expected to depend on time as well as on $x$ and $y$.
	To emphasise this point we rewrite (\ref{Hamil1}) in the form
	\begin{IEEEeqnarray}{rCl}
		k_x &= & \lambda\left(\frac{\partial \psi}{\partial x} \left.\frac{\partial S_{\!\perp}}{\partial \psi}\right |_0+ \left(\omega\frac{\partial y}{\partial \psi}-\frac{d\omega}{d\psi}t \frac{\partial \psi}{\partial x} \right) \frac{\partial S_{\!\perp}}{\partial \phi}\right) \nonumber \\
		k_y &= & \lambda \left(\frac{\partial \psi}{\partial y}\left. \frac{\partial S_{\!\perp}}{\partial \psi}\right |_0-\left(\omega\frac{\partial x}{\partial \psi} +\frac {d\omega}{d\psi}t \frac{\partial \psi}{\partial y} \right) \frac{\partial S_{\!\perp}}{\partial \phi}\right).\label{Hamil3}
	 \end{IEEEeqnarray}
	Here $|_0$ denotes that a derivative is to be taken ignoring the $\psi$ dependence of $\omega$, which is now taken into account by the terms with a factor $t.$ In the limit $t \rightarrow \infty$ we have 
	\begin{equation}
		k_x^2+k_y^2 \sim \lambda^2 \left(\frac{d\omega}{d\psi}\right) ^2 \left(\frac{\partial S_{\!\perp}}{\partial \phi}\right)^2 |{\bf v}|^2 t^2. \label{kinc}
	\end{equation}
	The right hand side of the above is the product of $|{\bf v}|^2t^2$ and a factor that is constant on a streamline indicating that the magnitude of the wavenumber increases to arbitrarily large values at all points on it.

\subsubsection{The special case of $n=\pm 1$ for a Kida vortex}\label{specialkida}
	For a Kida vortex in a Keplerian background, $\omega$ is constant so that terms $\propto t$ in (\ref{Hamil3}) are absent. We also note that only terms with $n=\pm 1$ are present in the representations given by (\ref{Fx}) and (\ref{Fy}) such that the particle trajectories on streamlines are given by 
	\begin{IEEEeqnarray}{rCl}
		x	&=&	C_K\psi^{1/2}\cos(\phi)\quad,\nonumber \\
		y	&=& -C_K\chi\psi^{1/2}\sin(\phi), \label{eq:part_traj}
	\end{IEEEeqnarray}
	with the amplitude factor $C_K$ being given by
	\begin{equation}
		C_K=(4(\chi-1)/(3\Omega_P\chi))^{1/2}.
	\end{equation}
	We now adopt $S_{\!\perp}= -S_K\psi^{1/2}\cos(\phi+\phi_0-\omega t),$ where $S_K$ and $\phi_0={\omega{\overline t} -\pi/2}$ are constants, and recall that for the Kida vortex we have ${\omega=3\Omega_P/(2(\chi-1))}$. The wavenumber is found from ${\bf k}=\lambda\nabla S_{\!A}$. With the help of (\ref{Hamil3}), we find that ${\bf k}$ is indeed given by (\ref{solution_k}) provided that we identify $k_z = k_0\cos(\theta)$ and $\lambda \omega C_K S_K/2 = k_0\sin(\theta)$.

	We confirm that although this time-dependent wavenumber was derived from terms with $n=\pm 1$ and the time-independent form is derived adopting $n=0$, one obtains equations governing the stability of a Kida vortex that are independent of   which is chosen. This is because, for the Kida vortex, the equations are invariant to a shift in the origin of time on a streamline and thus independent of ${\overline t}.$ This means that we may specify
	 $(k_x, k_y) \propto \nabla \psi$ in either case.

	However, it is important to note that in the generic case for which $\omega$ is not the same on different streamlines, one must adopt the time-independent form if modes growing exponentially with time in the usually expected manner are to be obtained. If a wavenumber increases linearly with time only temporary exponential growth is expected \citep[eg.][]{Goldreich1965}, with perturbations ultimately subject to at most power law growth with time thus requiring a nonlinear analysis to determine the outcome. This can be shown to be the case for the systems considered here (see below). Accordingly we extend the linear stability analysis for the Kida vortex to more general cases by adopting the time-independent wavenumber $(n=0)$, reserving use of the form given by (\ref{solution_k}) only for cases for which $\omega$ is constant.

	We comment that the situation here is analogous to the one that occurs for a differentially rotating disc for which fluid elements orbit on circles. The time-independent wavenumber modes here correspond to axisymmetric modes there. The modes with wavenumbers that increase with time correspond to non axisymmetric modes in the disc case. 

\subsection{Vertical stabilty}\label{vertical_stab}
	We now discuss vertical stability for which $k_z=0$ and the vertical velocity perturbation is zero. From the above discussion we expect that ultimately $k_z^2/(k_x^2+k_y^2) \rightarrow 0$ for $t \rightarrow \infty$ for choices of wavenumber that ultimately increase linearly with time. Accordingly, in this limit the discussion of vertical stability given here should apply. We adopt the Eulerian formulation leading to the linear equations (\ref{Eulerianp}). 
	The linearized incompressibility condition gives ${\bf k}\cdot {\bf v}'=0$. Thus we may set $v'_x=\eta k_y,$ and $v_y'=- \eta k_x$ for some scalar $\eta$.
	The linearized form of the condition that $\rho$ is fixed on fluid elements gives
	\begin{equation}
		\frac{D\rho'}{Dt}= -{\bf v}'\cdot \nabla \rho, \;\, \text{or equivalently} \;\, \frac{D\rho'}{Dt}= \eta {\bf k}\cdot{\bf v}\frac {d\rho}{d\psi}. \label{linrho}
	\end{equation}
	Eliminating $P'$ from the $x$ and $y$ components of the first of the equations (\ref{Eulerianp}) and making use of (\ref{kdot}) we obtain a relation between $\eta$ and $\rho'$ in the form
	\begin{equation}
		\frac{D (\eta |{\bf k}|^2)}{ D t} = - \frac{\rho'}{\rho^2}({\bf k}\times \nabla P)\cdot {\hat{\bf k}}. \label{2emp} 
	\end{equation}
	Equations (\ref{linrho}) and (\ref{2emp}) provide a pair of first order ordinary differential equations for which the integration is taken around streamlines. We note that in the general case from (\ref{Hamil3}) it is seen that ${\bf k}$ is not a periodic function of time and so they do not lead to a Floquet problem.

\subsubsection{Linear stability for the general vortex with shear}
	To discuss stability further set $\eta = W/|{\bf k}|^{2}$ and a new scaled time variable $\tau$, defined through $d\tau= |{\bf k}|^{3/2}dt$, to get the pair of equations
	\begin{IEEEeqnarray}{rCl}
		\frac{D\rho'}{D\tau} &=& \frac{ W{\bf k}\cdot{\bf v}}{ |{\bf k}|^{7/2}} \frac {d\rho}{d\psi},\label{linrho2} \\
		\frac{D W}{ D \tau} &=& - \frac{\rho'}{\rho^2}\frac{({\bf k}\times \nabla P)\cdot {\hat{\bf k}}}{|{\bf k}|^{3/2}}. \label{3emp}
	\end{IEEEeqnarray}
	In this form, provided $\partial S_{\!\perp}/ \partial \phi \ne 0,$ the coefficients of $W$ in (\ref{linrho2}) and $\rho'$ in (\ref{3emp}) tend to zero for large $\tau$ in such a way that there can be no exponentially growing solutions that apply at large times (although weaker growth could occur). Note in this context that although ${\bf k}$ increases linearly with time, equation (\ref{Hamil3}) implies that ${\bf k}\cdot {\bf v}$ remains bounded. 

\subsubsection{Vortex with Kida streamlines and density gradient in a Keplerian background}\label{vertparam}
	We now consider the situation for a vortex which is assumed to have stream function given by $(\ref{Kidasol})$ and non-constant $\rho= \rho(\psi)$, namely the polytropic model considered in Section \ref{polyeq}. In this case the pressure $P=P(\psi)$ and $d\omega(\psi)/d\psi =0$. On account of the latter relation we can adopt the solution given by (\ref{solution_k}) for $\theta =\pi/2,$ which corresponds to $k_z=0$, without encountering problems of the wavenumber increasing linearly with time. The coordinates on a streamline can be specified as indicated in Section \ref{specialkida}. Then equations (\ref{linrho}) and (\ref{2emp}) can be combined to give a second order ordinary differential equation for $\rho'$ of the form 
	\begin{IEEEeqnarray}{rCl}
		\frac{D}{Dt}\left[\left(\chi^2+1+(\chi^2-1)\cos 2(\omega t-\phi_0)\right) \frac{D\rho'}{Dt}\right]=&	\nonumber	\\
		-\,\frac{6\Omega_P \chi \sin^2 (\omega t_0 -\phi_0)\,\psi}{(\chi-1)\,\rho^2}\frac{dP}{d\psi}&\frac{d\rho}{d\psi}\,\rho',	\label{Hill0}
	\end{IEEEeqnarray}
	Note that the terms on the right hand side of (\ref{Hill0}) multiplying $\rho'$ are constant on a streamline. Thus if we set 
	\begin{equation}
		{\cal Z} = \left(\, \chi^2+1 +(\chi^2-1)\cos2(\omega t-\phi_0)\right) \frac{D\rho'}{Dt},
	\end{equation}
	we find that ${\cal Z}$ satisfies a form of Hill's equation that can be written in the form
	\begin{equation}
		\frac{D ^2 {\cal Z} }{ D t^2}= -\,\frac{6 \Omega_P\chi\sin^2 (\omega t_0 -\phi_0)\,\psi}{(\chi-1)\bigg(\, \chi^2+1 +(\chi^2-1)\cos2(\omega t-\phi_0)\bigg)\,\rho^2}\frac{dP}{d\psi}\frac{d\rho}{d\psi}{\cal Z}.  \label{Hill}
	\end{equation}
	This equation can be interpreted as describing the evolution of a gravity wave with a time-dependent wavenumber. It can rewritten in the form
	\begin{equation}
		\frac{D ^2 {\cal Z} }{ D t^2}= -\frac{q}{1 +\chi^2 + (\chi^2-1)\cos2(\omega t-\phi_0)}{\cal Z}, \label{Hill1}
	\end{equation}
	where $q$ is a constant that can be scaled up to a maximum value $q_{max}= 6 \Omega_P\chi (\chi-1)^{-1} \psi \rho^{-2}(dP/d\psi) (d\rho/d\psi)$ by adjusting the value of $\sin^2 (\omega t_0-\phi_0)$ through the specification of $\omega t_0 -\phi_0$. The quantity $q_{max}$ can be interpreted as the square of a buoyancy frequency. The possibility of parametric instability is expected when this frequency is large enough to be comparable to the vortex frequency $\omega$. Solutions of this are discussed below.

\section{Numerical procedures and numerical results}\label{sec:numerical_method}
	
	\begin{figure*}
		\begin{minipage}{177mm}
			\begin{center}
			    \subfigure[Kida vortex\label{Kidaandpol_kida}]
			    {
				    \includegraphics[width=0.45\textwidth]{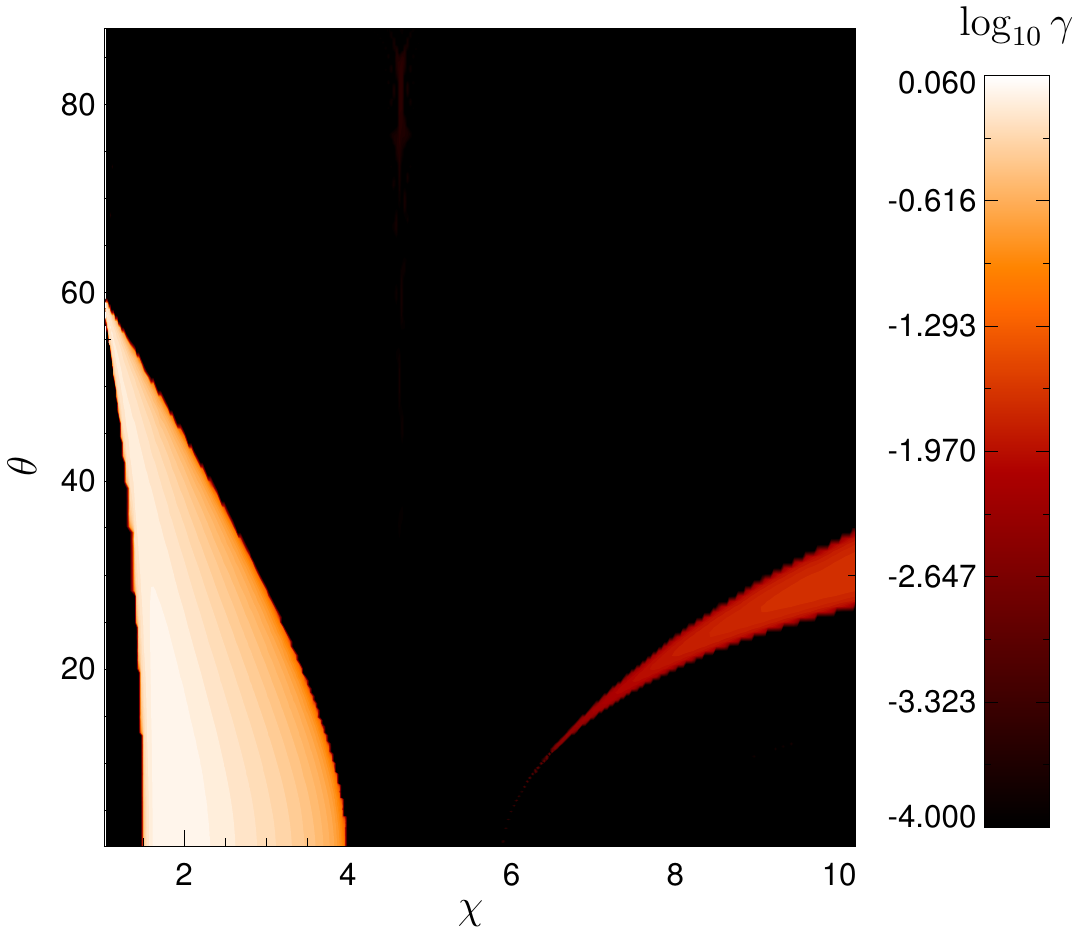}
				}
			    \begin{subfigure}[Polytropic model\label{Kidaandpol_pol}]
			    {
			    \includegraphics[width=0.45\textwidth]{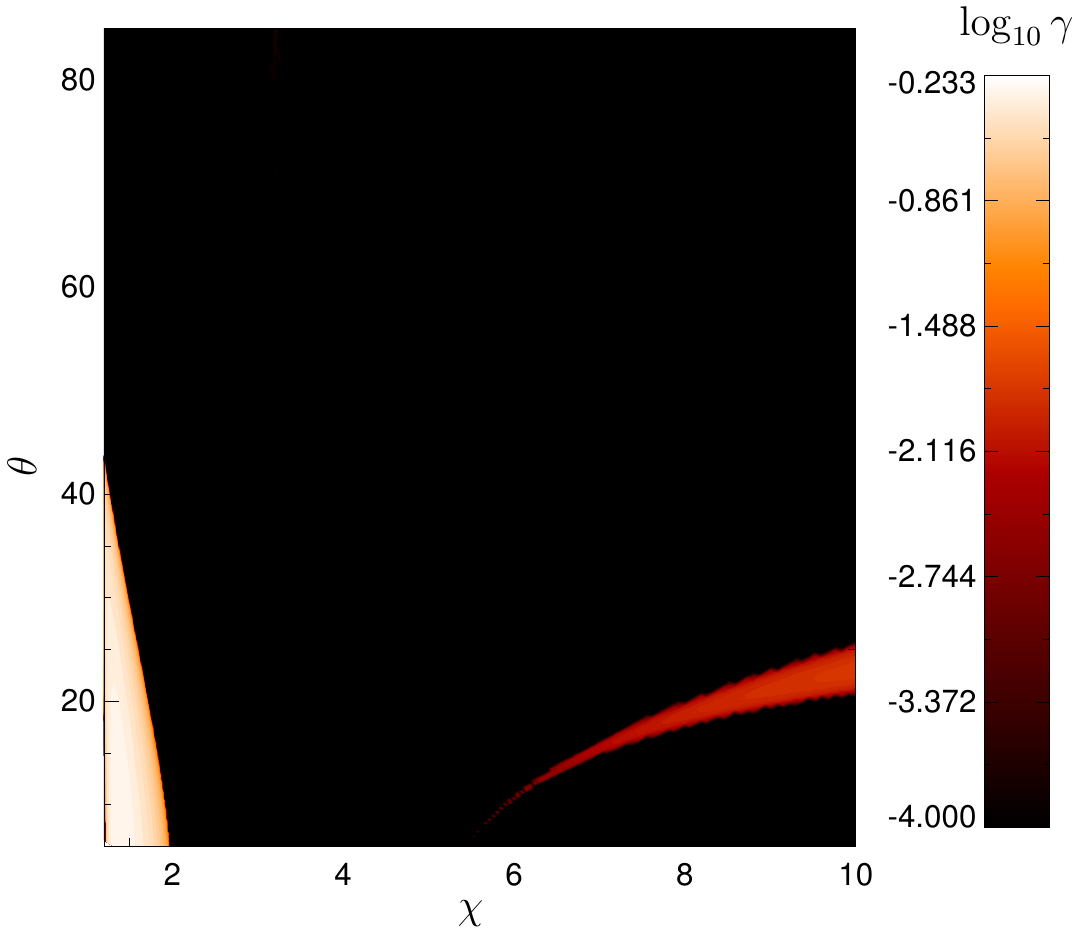}
			    }
			    \end{subfigure}
				\caption{ The stability of the Kida vortex (a) and polytropic model with uniform density in a non Keplerian background (b).}
				\label{Kidaandpol}
			\end{center}
		\end{minipage}
	\end{figure*}

	\begin{figure*}
		\begin{minipage}{177mm}
			\begin{center}
				\subfigure[$ \left\{\alpha,\beta,\rho_m\right\}= \left\{0.25,0,0\right\}$, with $y_{max}=0.5$.]
					{
					\includegraphics[width=0.45\textwidth]{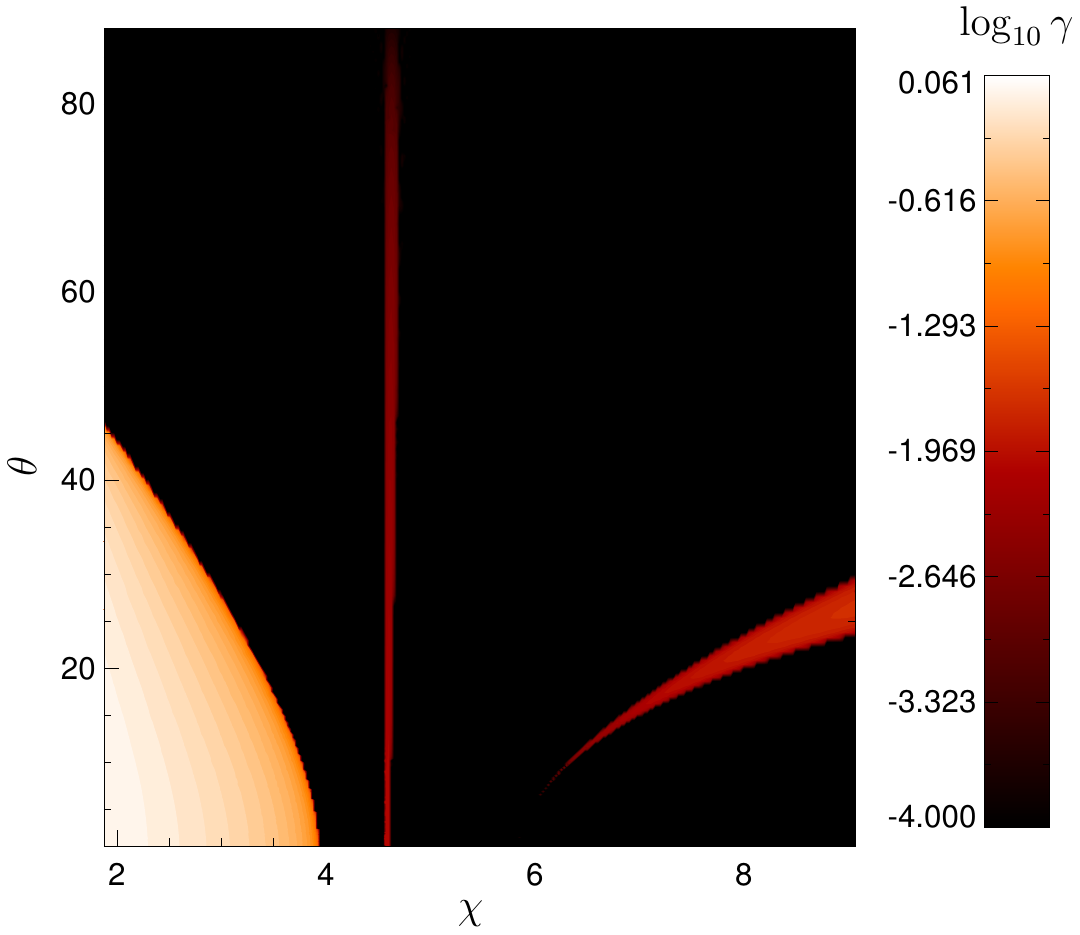}
					}
				\subfigure[$\left\{\alpha,\beta,\rho_m\right\}=\left\{0.25.0,0\right\}$, with $y_{max}=0.85$.]
					{
					\includegraphics[width=0.45\textwidth]{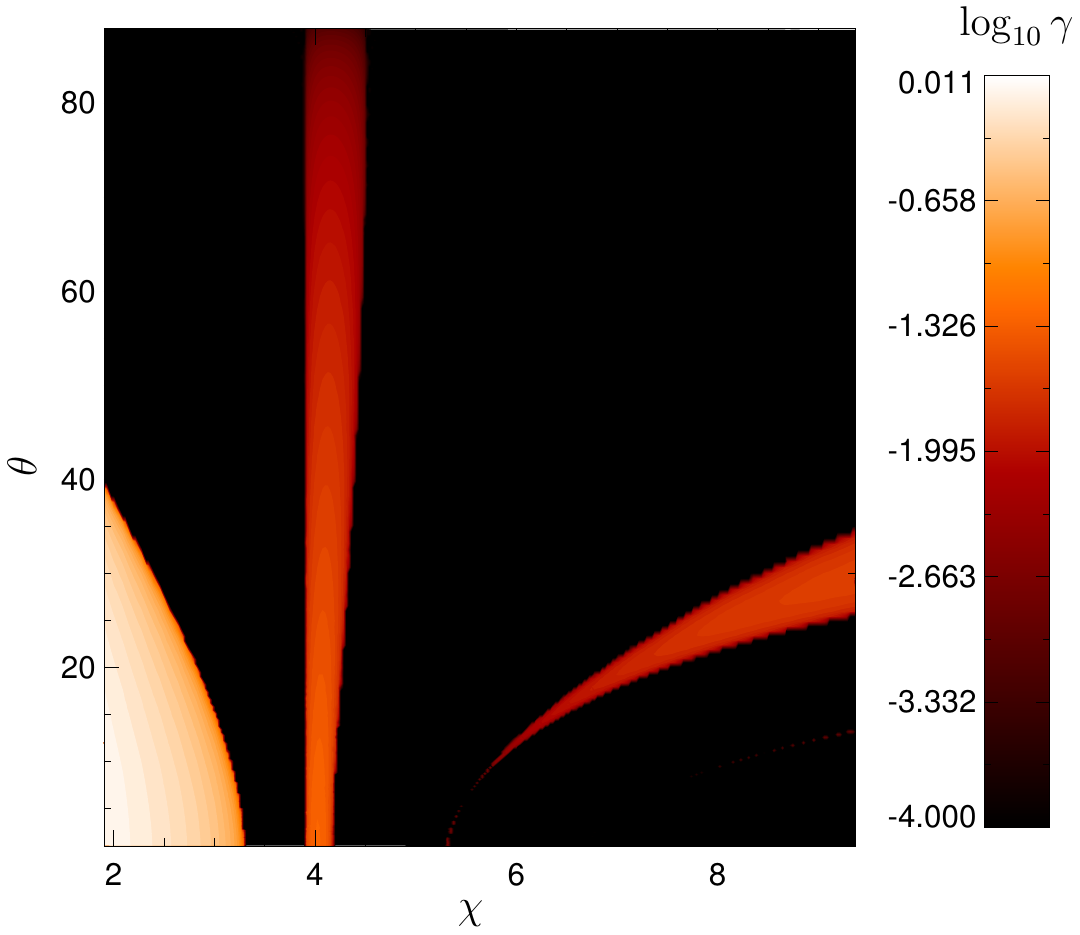}
					}
				\caption{The stability of motion in vortices with $\alpha=0.25$, on different streamlines. The parameter $\omega_m$ is varied to produce the range of aspect ratios shown for this streamline.}
				\label{stabilitynodensity}
			\end{center}
		\end{minipage}
	\end{figure*}	

\subsection{Steady states}
	We perform calculations to obtain steady state solutions for $\psi$ for vortices on a $512\times1024$ grid that covers the vortex core with $x$ and $y$ occupying the interval $[-1,1]$. This grid was chosen to ensure enough resolution to undertake the local stability analysis described below. However, very similar results were obtained when the grid resolution was reduced by a factor of two. After the solution inside is obtained inside the core, it can be readily utilised in order to determine the streamlines outside the core.

	We find that it is numerically convenient to gradually build up a density enhancement in the centre of the vortex. Thus we initially solve a reduced version of the Poisson equation with $B=0$ that is applicable to the uniform density case, namely
	\begin{equation}
		\nabla^2 \psi_1 = \mathcal{A}(\psi). \label{eq:numerics1}
	\end{equation}
	We start by defining the Bernoulli source term $\mathcal{A}$ to be a non-zero constant inside the unit circle passing through $(0,1)$ and zero outside this. We then solve equation (\ref{eq:numerics1}) by utilising the 2D Green's Function, obtaining
	\begin{equation}
		\psi_1(\mathbf{r})=\frac{1}{2\pi}\iint \log \vert \mathbf{r}-\mathbf{r'}\vert\, \mathcal{A}(\psi) \, d^2 \mathbf{r'}
	\end{equation}
	We then add in the streamfunction for the background shear, ${\psi_0=\frac{3}{4}\Omega_Px^2}$, to get the total streamfunction $\psi$. Setting the bounding streamline of the vortex core to pass through $(0,1),$ we then recalculate the boundary, rescale the Bernoulli source function as indicated below, and apply the Green's Function again to obtain an updated solution for $\psi_1$. 

\subsubsection{Rescaling}
	We define $\pi\omega_m=\int \mathcal{A}(\psi)\,dS,$ where the integral is taken over the total area inside the vortex core. This quantity is the magnitude of the circulation around the vortex contributed by the non background source. In order to rescale the Bernoulli source, we recalculated $\mathcal{A},$ adopting the required functional form, after every iteration. We then renormalized it, so as to ensure that $\pi\omega_m$ remained fixed. This procedure was iterated until there was no discernible difference in successive streamfunction and pressure contours. For vortices with $\mathcal{B} =0,$ this required generally about 50 iterations. Comparison with the analytic solution for the Kida vortex case (see Section \ref{sec:kidavortex}) enabled our numerical procedure to be checked (see below). 

\subsubsection{Solutions incorporating a density enhancement}\label{soldenh}
	In order to obtain solutions with increased density in the central parts of the vortex core we have also found solutions of equation (\ref{eq:psi1}) with both $\mathcal{A}\ne 0,$ and $\mathcal{B}\ne 0$. For these cases the pressure distribution has to be calculated at each iteration. This is found from equation (\ref{Pcalc}) which yields
	\begin{equation}
		\frac{P}{\rho}=\frac{3}{2}\Omega_P x^2 - \frac{1}{2} \vert \nabla \psi \vert^2 -\frac{1}{2}\Omega_P\psi + \int_{\psi_{b}}^{\psi} \mathcal{A}(\psi')\,d\psi'. 
	\end{equation}
	Our numerical models are characterised by input values of $\alpha$ and $\beta$ that are required to implement equations (\ref{eq:A}) and (\ref{eq:RHO}) respectively. It is convenient to scale the mass per unit length added to the vortex by imposing a fixed value of $\rho_m=\int[\,\rho(\psi)-\rho_0]\,dS$ and, as before $\pi\omega_m=\int \mathcal{A}(\psi)\,dS$, which is a measure of the additional imposed circulation. In this case the constants (\ref{eq:A}) and (\ref{eq:RHO}) are rescaled after every iteration to ensure the required fixed values of $\rho_m$ and $\pi\omega_m$ are maintained. 

	Solutions are thus characterised by the parameters $\left\{\alpha,\beta,\rho_m,\omega_m\right\}$, with the last two quantities expressed in our dimensionless units. Convergence for models with constant density was straightforward, although resolution issues arise for a fixed grid when $\alpha$ becomes too large on account of peaking of the vorticity profile in the centre of the vortex. 

	For models with increased density inside the vortex core, convergence was more difficult and required a starting model close to the final one. In order to deal with this we began by imposing only small increases to either of, or both, $\beta$ and $\rho_m$ from their values appropriate to an existing solution. These changes moved them in the direction of our target parameters. This procedure was especially necessary for the cores of vortices with non-uniform vorticity profiles. Having done this, a new form of $\psi$ was obtained as above by iterating the Green's function solution $N\ge 10$ times, at which point additional small increments of the order of $1\%$ were made to $\beta$ and $\rho_m$ and the process repeated until the target values were attained. After that the Green's function solution could be iterated to convergence. 

\subsection{Stability calculations}\label{Stabcalc}
	The procedure we used was to solve the system of   equations (\ref{3sEem} ) together with equation   (\ref{3sEce})   for the Eulerian perturbations as an initial value problem. As the integration is around steady state streamlines, state quantities need to be specified on them. A particular streamline has to be specified by identifying an initial location eg. $(0,y_{max}).$ The location on it as the integration proceeds is then specified by solving the equations 
	\begin{IEEEeqnarray}{rCl}
		\frac{Dx}{Dt}&=&\frac{\partial\psi}{\partial y} \nonumber \\
		\frac{Dy}{Dt}&=&-\frac{\partial \psi}{\partial x}. \label{rdot}
	\end{IEEEeqnarray}
	The growth rate of any instability present was obtained by solving the additional equation
	\begin{equation}
		\frac{D \gamma_{_2}}{Dt}= \frac{1}{2}\ln|{\bf v}'|^2.
	\end{equation}
	For a system with growth rate $\gamma,$ we expect that ultimately $\gamma_{_2} \rightarrow \gamma t^2/2$. Accordingly we determined $\gamma$ by making a parabolic fit to $\gamma_{_2}$ at large times. We found that integrations running for $1000$ circulations around streamlines could detect growth rates down to $\sim 0.001\Omega_P$.

	In order to get a global view of stability we need to consider models over a large range of aspect ratio and $k_z$. For cases with time-independent wavenumber, 
the latter quantity is specified through the ratio $\sqrt{k_x^2+k_y^2}/|k_z|\equiv \tan\theta$ evaluated at $(0,y_{max})$ on the stream line. We remark that this quantity is $\propto |{\bf v}|$ on the streamline and so is
 minimised at the chosen location. Regions of instability often require high resolution in this phase space in order to be adequately monitored.
 Typically for a model with specified $\alpha,\beta$ and $\rho_m$, we required a $300\times 300$ grid in the $(\chi,\theta)$ plane (see results below). In order to perform these time-consuming calculations, we used the fact that all required quantities on a given streamline can be determined in terms of $\psi$. To facilitate this we made two dimensional least square polynomial fits of up to sixth order to $\psi$ inside the vortex core for a grid of calculated models. We note that this should give a good representation for streamlines close enough to the centre and the onset of parametric instability (see section \ref{Parc}). In practice we found that even fourth order polynomial fits gave results not significantly different from those presented below.

 	\begin{figure*}
		\begin{minipage}{177mm}
			\begin{center}
				\subfigure[$\left\{\alpha,\beta,\rho_m\right\} = \left\{0.5,0,0\right\}$]
					{
					\includegraphics[width=0.45\textwidth]{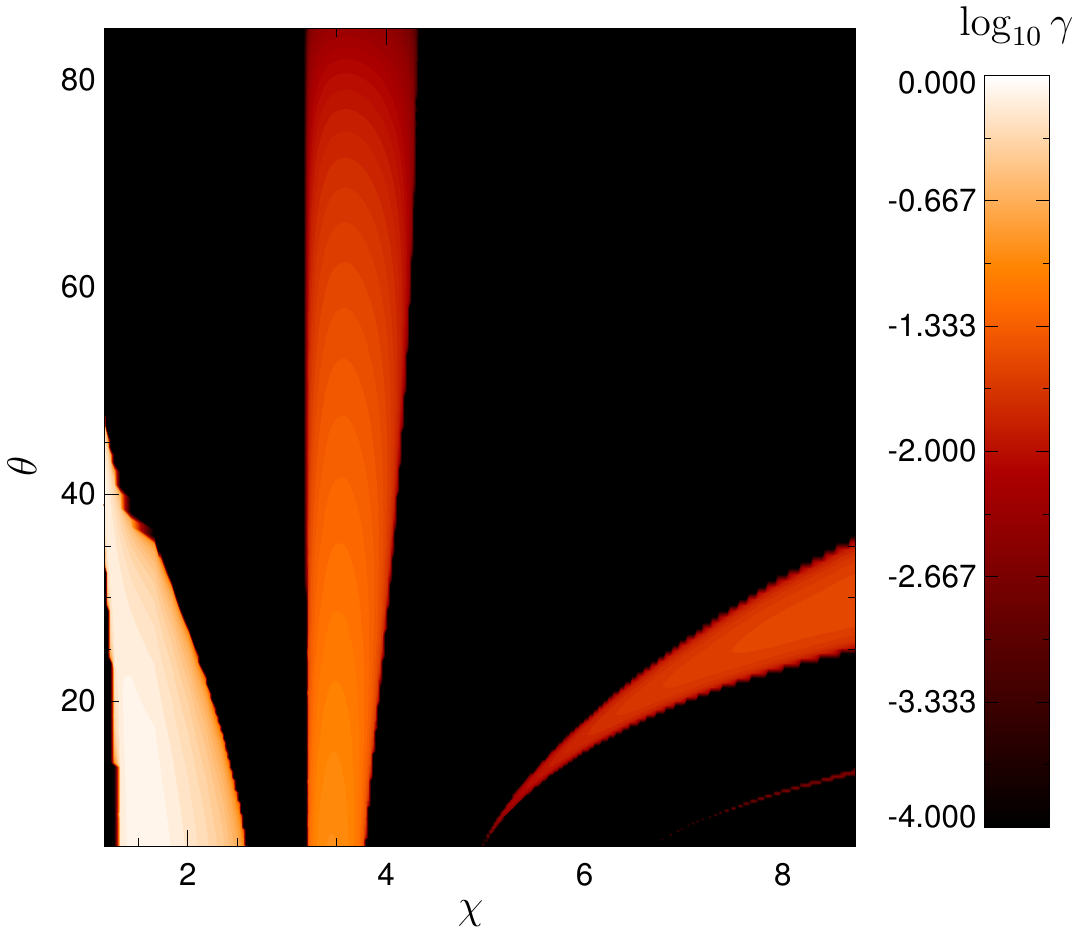}
					}
				\subfigure[$\left\{\alpha,\beta,\rho_m\right\} =\left\{1.0,0,0\right\}$]
					{
					\includegraphics[width=0.45\textwidth]{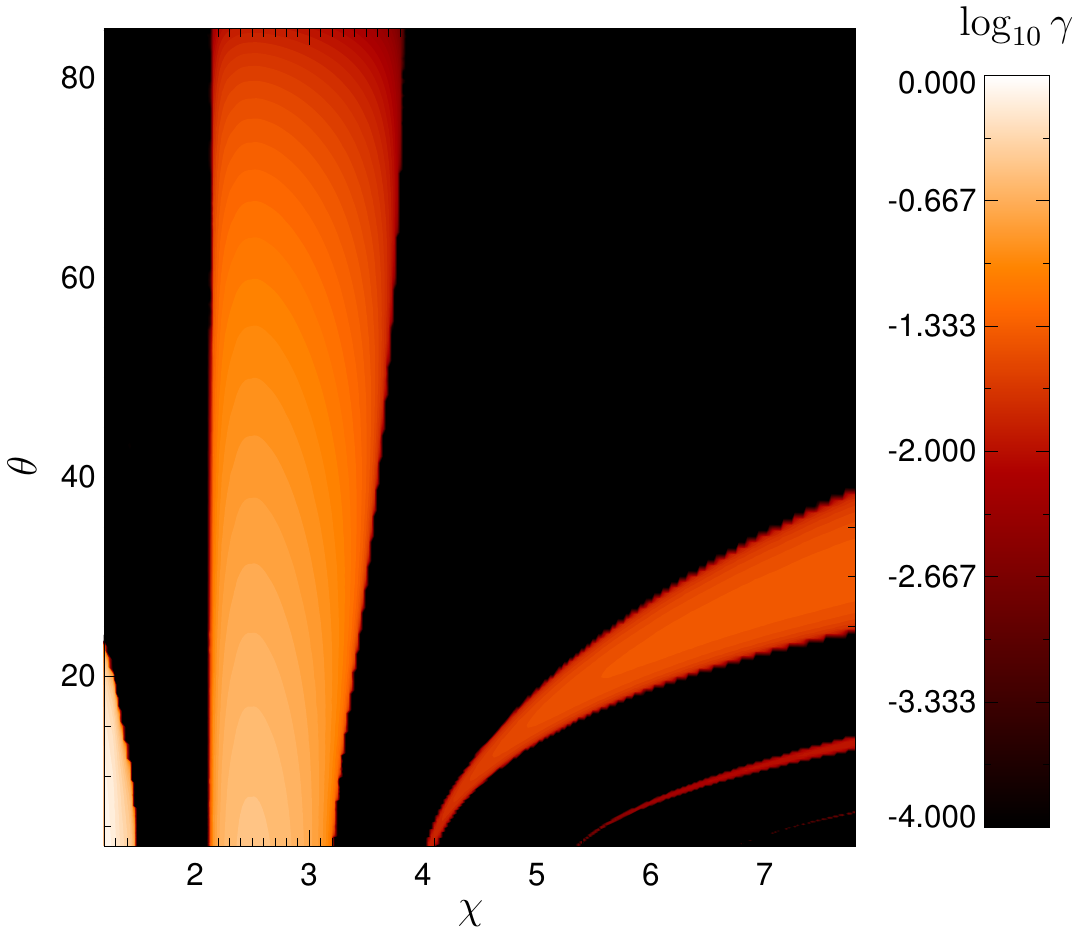}
					}
				\subfigure[$\left\{\alpha,\beta,\rho_m\right\} =\left\{2.0,0,0\right\}$]
					{
					\includegraphics[width=0.45\textwidth]{figs/stab1000}
					}
				\subfigure[Point vortex] 
					{
					\includegraphics[width=0.45\textwidth]{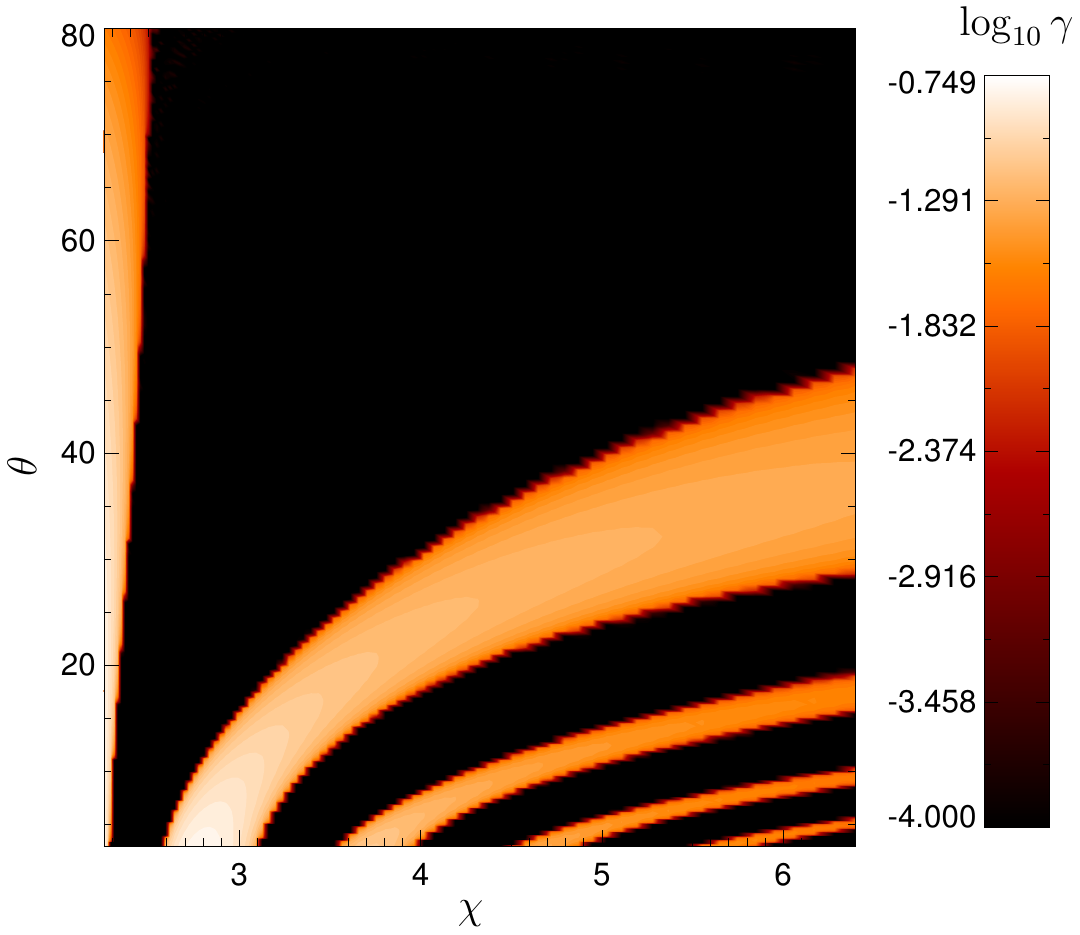}
					}
				\caption{As in Figure~\ref{stabilitynodensity} but for different values of $\alpha$, with $y_{max}=0.85$ as standard. In addition the stability of the point vortex model is shown. In this case the streamline aspect ratio is varied by changing the strength of the point vortex.}
				\label{stabilityvaryalphanodensity}
			\end{center}
		\end{minipage}
	\end{figure*}

\subsection{Numerical results} \label{sec:numerical_results}
\subsubsection{Steady state models}

	We show the streamlines for vortices with $\omega_m=0.09$ and density equal to the background value obtained using our numerical procedure in Figure~\ref{equilnodensity}. These include a Kida vortex with $\chi= 5$ and two vortices with non-uniform vorticity distributions in their cores, with $\alpha=1$ and $\alpha=2$. In Figure~\ref{equilnodensity} the vorticity distribution, the $\psi$ contours and the pressure distribution are shown. 

	In addition, the period $\tilde{P}$ required to circulate around a streamline is plotted against the value of the positive $y$-coordinate $y_{max}$, where it intersects the $y$-axis. The values of $\tilde{P}_{\text{\sc{kida}}}$ indicated are calculated analytically using the formula
	\begin{equation}
		\tilde{P}_{\text{\sc{kida}}}=\oint \frac{ds}{\vert\nabla\psi\vert}=\frac{4\pi}{3}(\chi-1),
	\end{equation}
	which is independent of the streamline chosen (the Kida solution has constant period throughout the vortex patch). The results for the Kida vortex are in good agreement with analytic expectation. Note that the periods we plot were obtained by locating coordinate extrema from the results of numerical integrations of fluid particles moving around streamlines stored on a coarse temporal grid. This results in some low-level jitter at a relative level $\sim 10^{-3}$. This is also a measure of the departure from a constant value of the period obtained numerically in the Kida vortex case.
	 
	Note too that a significant non-constant value of the period to circulate around a streamline indicates the presence of shear inside the vortex. The latter becomes more noticeable as $\alpha$ increases. Apart from in the case of the vortex illustrated in the bottom panels, there is a pressure maximum at the centre and in the latter case there is a saddle point. We find that the transition from central saddle point to maximum occurs at smaller aspect ratios as $\alpha$ increases and the vorticity distribution becomes more centrally concentrated. 
	This shifts the central saddle point instability described above to central values of $\chi < 4$. We recall that cases for which the pressure distribution has a maximum at the centre of the vortex are of interest as they are expected to attract dust \citep{Whipple1972, Cardoso1996}. In addition, streamlines for vortices with centrally concentrated vorticity sources tend to become pinched towards the $y$-axis as compared to the Kida case as one moves out from the centre. This leads to a relative increase in the circulation period.

	Vortices with varying density excess [for which $\omega_m=0.09$, $\alpha=0$ and $\beta=1$] are similarly illustrated in Figure~\ref{equilKidadensity}. In these cases because the Bernoulli source is uniform, the density distribution is shown. For these models the central density ranges between $1.3$ and $2.5$ times the background level. The pressure has a central maximum in all cases. A small amount of internal shear is present $\sim 0.001\Omega_P$. However, this could be significant for linear perturbations (see below).

	Finally vortices with a non-uniform Bernoulli vorticity source, with $\alpha = 0.5$, $\omega_m=0.09$, and non-zero density enhancement parameter $\rho_m$ are illustrated in Figure~\ref{equilnonKidadensity}. The vorticity profiles in these vortices are non-uniform, resulting in significant variation of the period and hence significant internal shear.

    \begin{figure*}
        \begin{minipage}{177mm}
            \begin{center}
                \subfigure[$\left\{\alpha,\beta,\rho_m\right\} =\left\{0,1,0.5\right\}$, streamline passing through $(0.0,0.85)$.]
                {
				\includegraphics[width=0.45\textwidth]{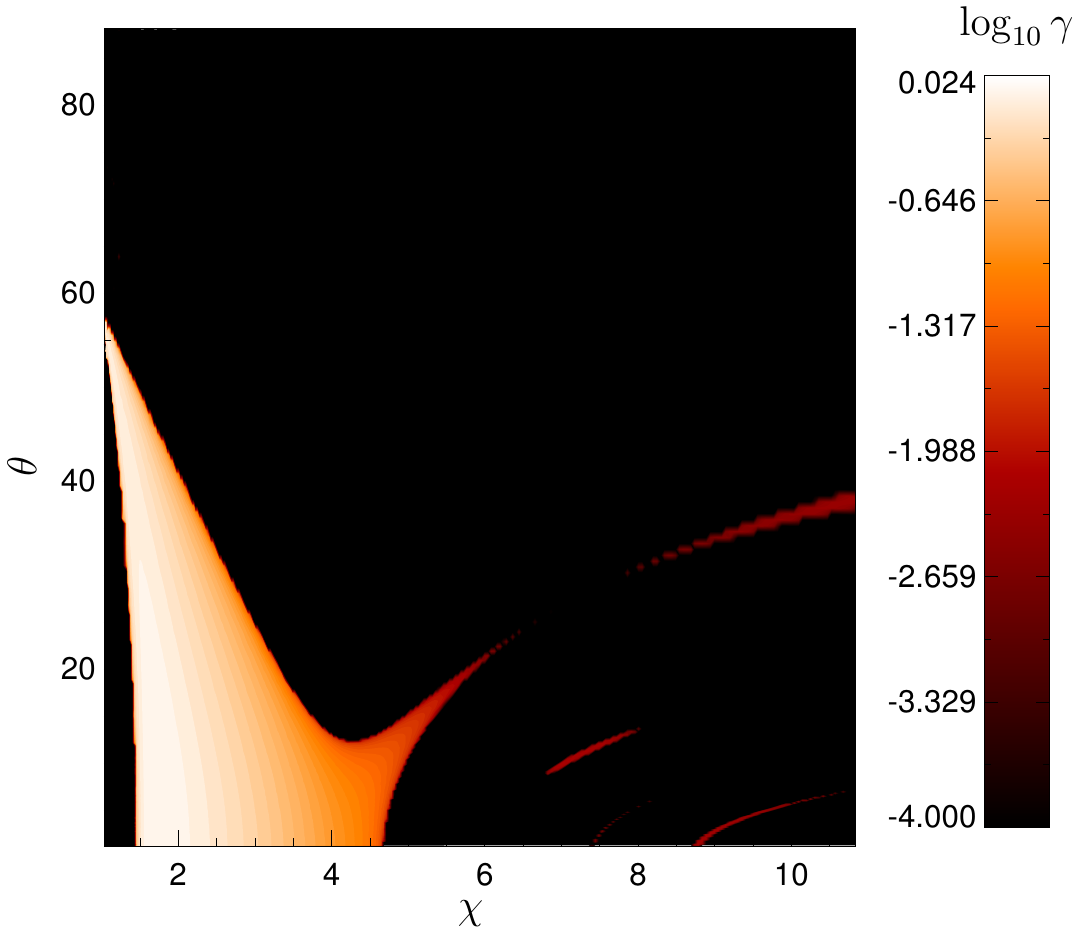}
                }
	            \subfigure[$\left\{\alpha,\beta,\rho_m\right\} =\left\{0,1,2\right\}$, streamline passing through $(0,0.5)$.]
                {
				\includegraphics[width=0.45\textwidth]{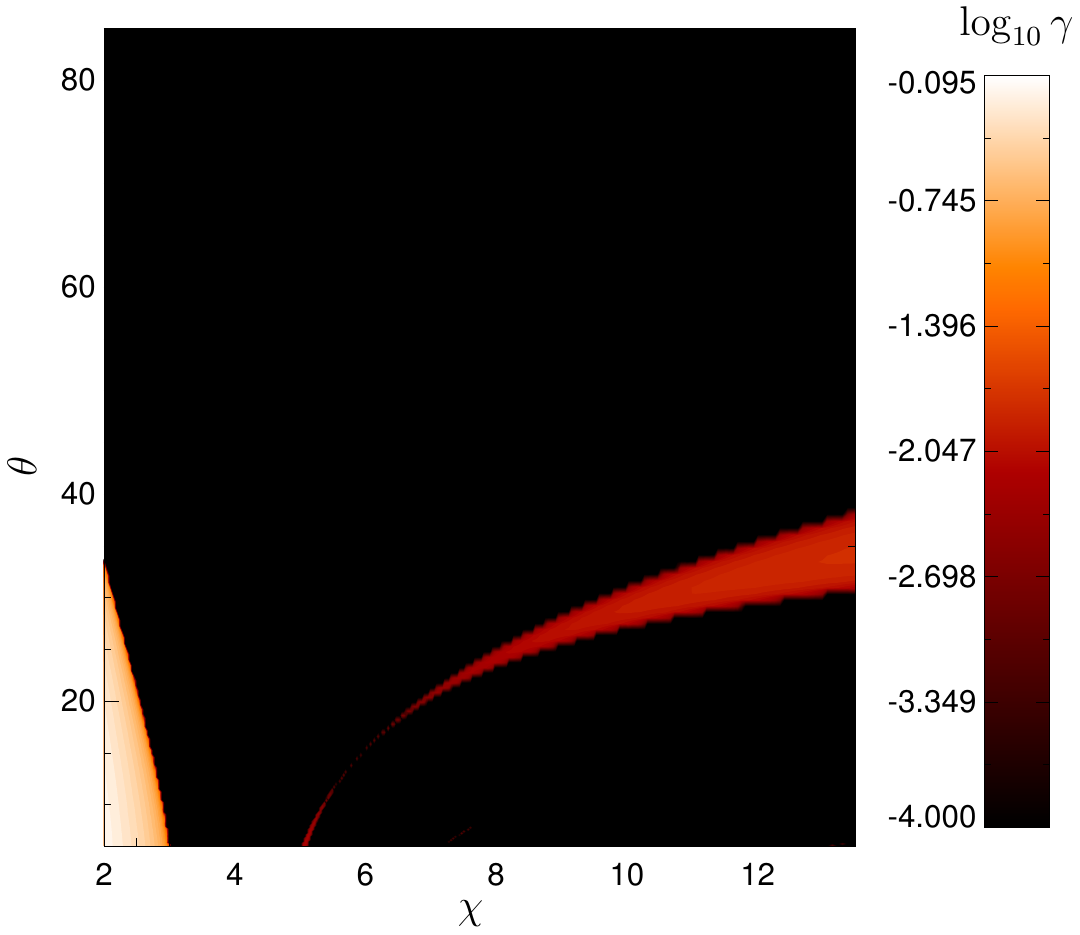}
				}	
	            \subfigure[$\left\{\alpha,\beta,\rho_m\right\} =\left\{0,1,2\right\}$, streamline passing through $(0,0.67)$.]
                {
				\includegraphics[width=0.45\textwidth]{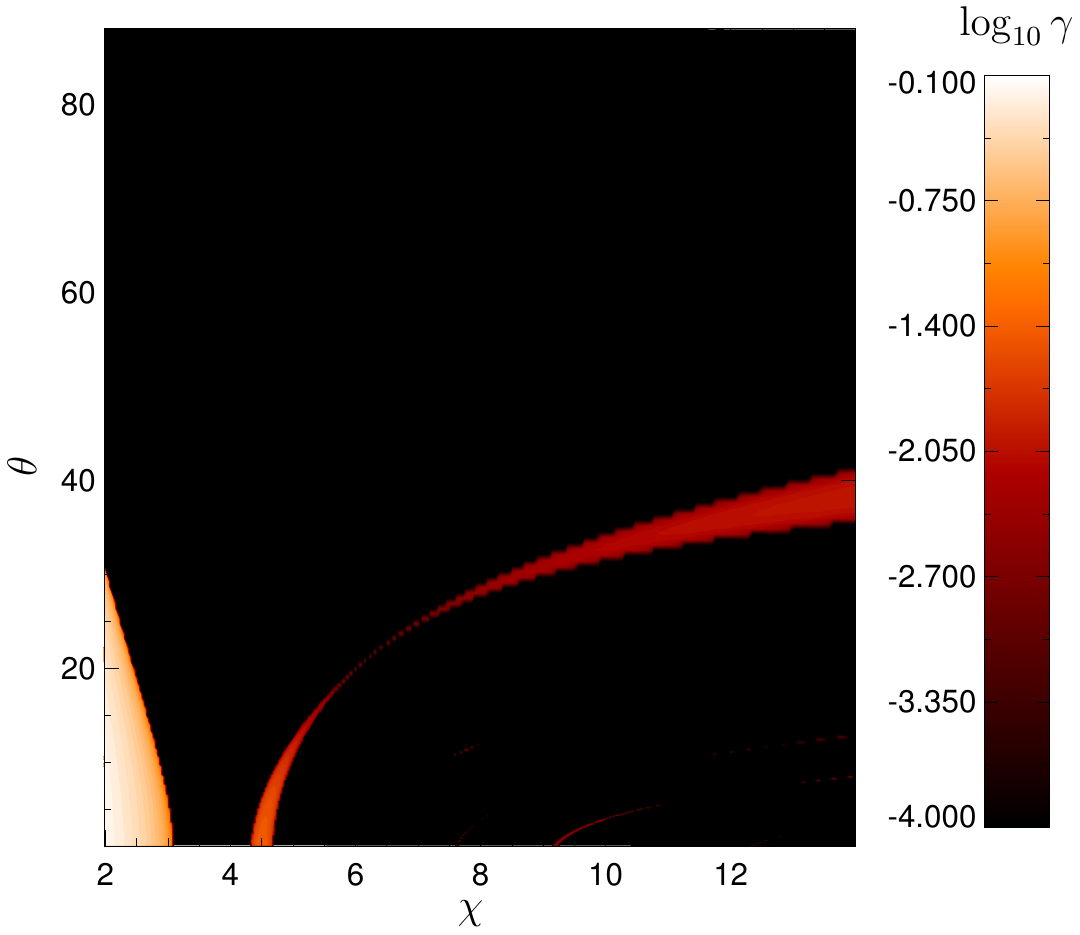}
				}
	            \subfigure[$\left\{\alpha,\beta,\rho_m\right\} =\left\{0,1,2\right\}$, streamline passing through $(0,0.85)$.]
                {
				\includegraphics[width=0.45\textwidth]{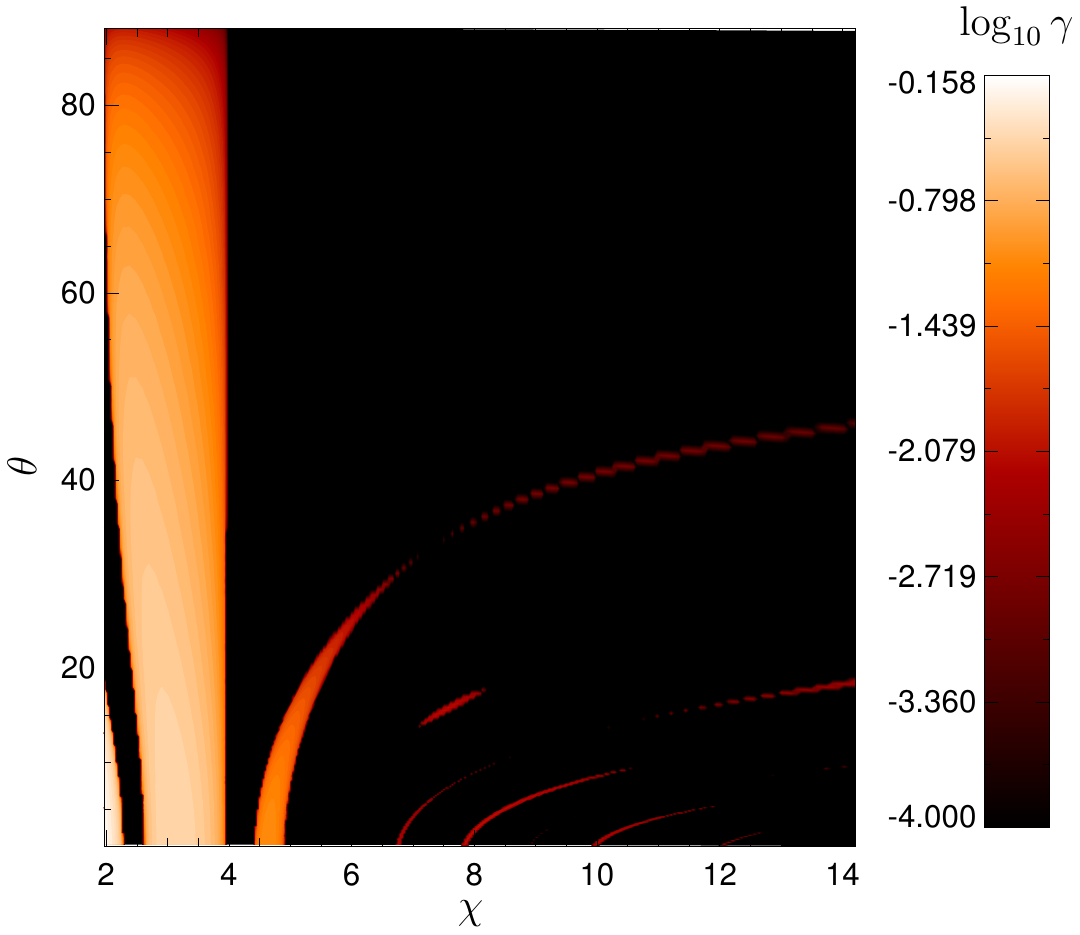}
				}
	            \caption{The stability vortices with $\alpha=0,$ $\beta=1$ and varying $\rho_m$. Results are shown for streamlines with $y_{max}=\{0.5,\,0.67,\,0.85\}$. The parameter $\omega_m$ is varied to produce the range of aspect ratios shown.}
	            \label{stabilityKidadensity}
	    \end{center}
	\end{minipage}
	\end{figure*}

	\begin{figure*}
        \begin{minipage}{177mm}
            \begin{center}
                \subfigure[$\left\{\alpha,\beta,\rho_m\right\} =\left\{0.25,1,0.5\right\}$]
                {
				\includegraphics[width=0.45\textwidth]{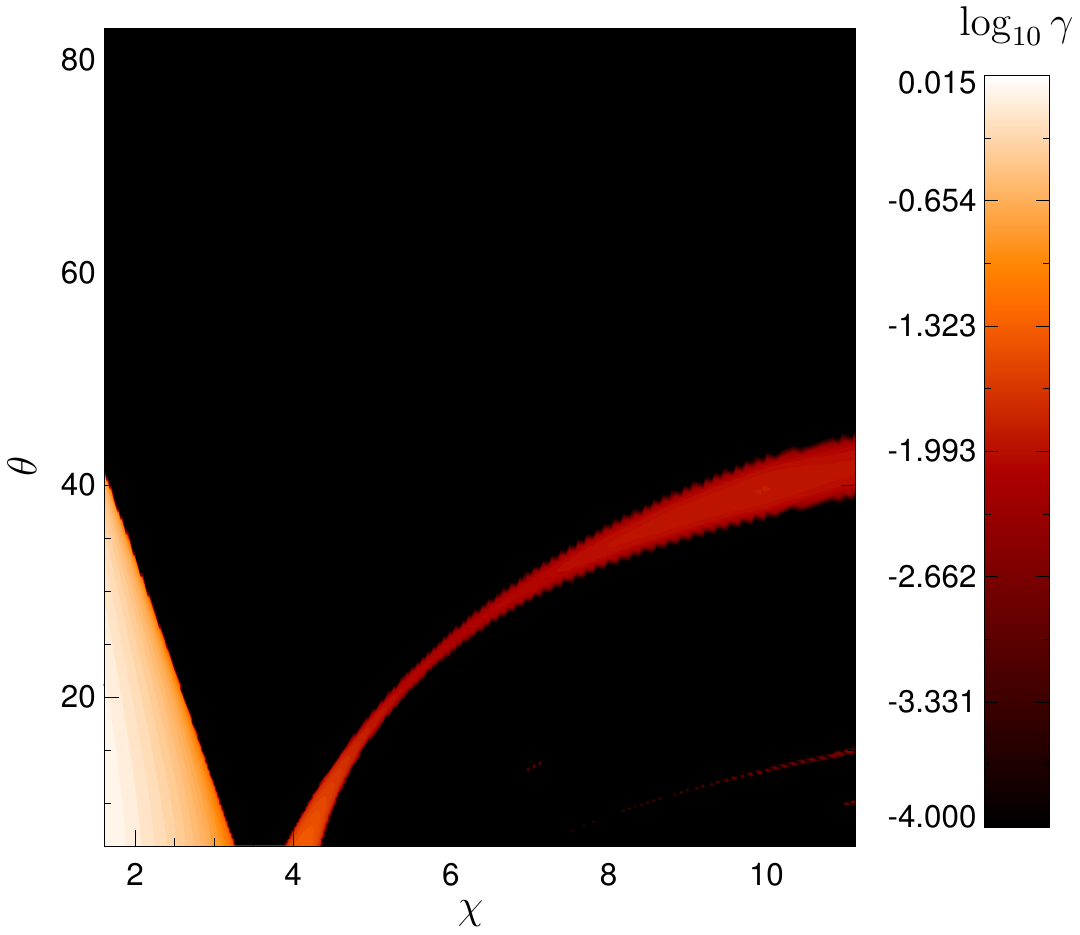}
                }
                \subfigure[$\left\{\alpha,\beta,\rho_m\right\} =\left\{0.5,1,0.3\right\}$]
                {
				\includegraphics[width=0.45\textwidth]{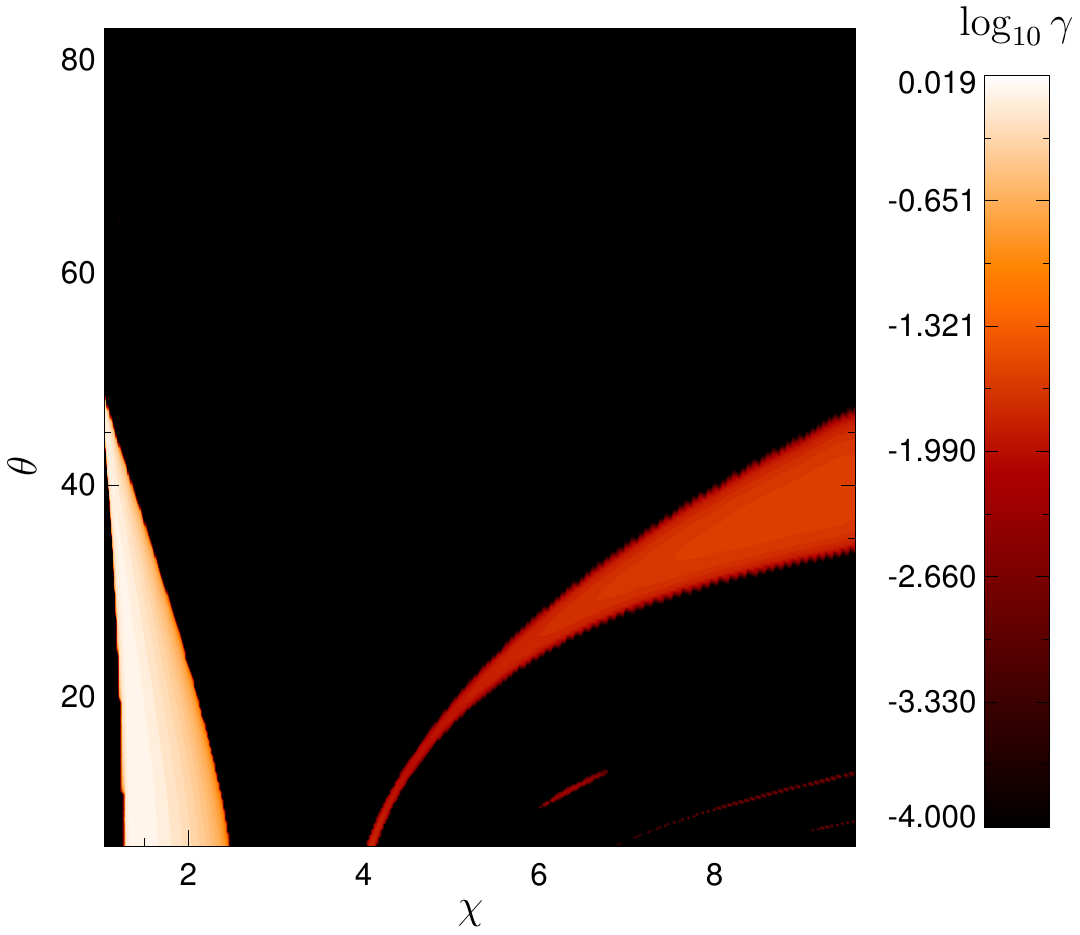}
				}
                \caption{The stability of vortices with $\alpha=0.5$ and non uniform density. The results shown are for the streamline that passes through $(0.0,0.85)$.}
                \label{stabilitynonKidadensity}
            \end{center}
	    \end{minipage}
    \end{figure*}

\subsubsection{Stability calculations}\label{satbcalc}

	The stability of both Kida vortices and the polytropic model with uniform density in a non-Keplerian background (see Section \ref{polyeq}), determined using our procedures as viewed in the $(\chi, \theta)$ plane are illustrated in Figures~\ref{Kidaandpol}(a) and (b) respectively. The results for the Kida vortices are in good agreement with previous work \citep[see][]{Lesur2009}. The results for the polytropic model are qualitatively similar but with the region of strong saddle point instability shifted to smaller values of $\chi < 2$.
	  	  
	In order to study the effects of introducing a variable vorticity profile in the core, we illustrate the stability of vortices with $\alpha=0.25$ in the $(\chi,\theta)$ plane in Figure~\ref{stabilitynodensity}. The stability of motion on streamlines with $y_{max}=0.5$ and $ y_{max}=0.85$ (Figures~\ref{stabilitynodensity}(a) and (b) respectively) is shown. The parameter $\omega_m$ is varied to produce the range of aspect ratios shown for this streamline. As indicated in Section \ref{Parc}, moving out from the vortex centre we expect parametric instability to occur for $\chi \sim 4.65$. This is visible for the streamline with $y_{max}=0.5$ for which a narrow vertical instability band appears at this location. This band is seen to broaden while the instability region at large $\chi$ extends towards the $\chi$-axis at smaller values of $\chi$. In addition the region associated with saddle point instability shifts to smaller values of $\chi$.

	In Figure~\ref{stabilityvaryalphanodensity} we show similar results to those presented in Figure~\ref{stabilitynodensity} for streamlines with $y_{max}=0.85$ and $\alpha = \{0.5,\,1.0,\,2.0\}$. As $\alpha$ increases, the instability band originating from $\chi=4.65$ widens and moves to smaller values of $\chi$, while the small $\chi$ region associated with saddle point instability eventually disappears. Several additional instability bands appear at larger values of $\chi$. We also considered a vortex point source model. This adopts the streamfunction $\psi = K\ln|{\bf r}| +3\Omega^2x^2/4$ which corresponds to a Bernoulli source localised at the centre. It can be viewed as the limiting case of large $\alpha$. The aspect ratio of the streamline for a fixed $y_{max}=0.85$ is fixed by an appropriate choice of the constant $K$. The stability properties are obtained using the same procedures as for the other models. Results are given in Figure~\ref{stabilityvaryalphanodensity}. The behaviour is very similar to that seen for the case with $\alpha=2$, with the $(\chi, \theta)$ plane filling up with instability bands. Growthrates at larger values of $\chi \gtrsim 5$ are $\sim 0.05\Omega_P$. We comment that although these models do not have any density excess, because they are relevant to streamlines outside a high density core they are of generic significance for vortices accumulating dust in their cores.

\begin{figure}
            \includegraphics[width=0.45\textwidth]{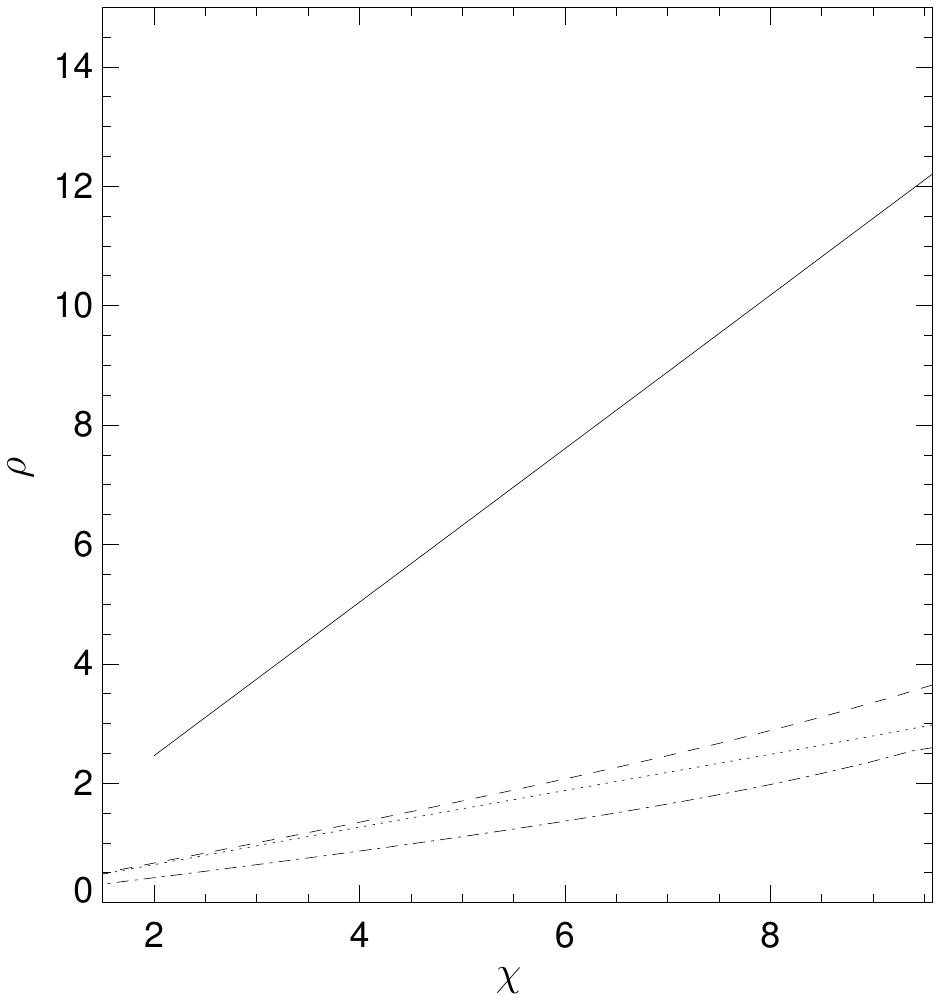}
            \caption{The density excess above the background measured in units of the background density for the vortices with $\left\{\alpha,\beta,\rho_m\right\} =\left\{0,1,2\right\}$ (solid line), $\left\{\alpha,\beta,\rho_m\right\} =\left\{0,1,0.5\right\}$ (dotted line), $\left\{\alpha,\beta,\rho_m\right\} =\left\{0.25,1,0.5\right\}$(dashed line) and $\left\{\alpha,\beta,\rho_m\right\} =\left\{0.5,1,0.3\right\}$ (dashed-dot line).}
            \label{densitycurves}
    \end{figure}

	The stability of vortices with $\alpha=0$, $\beta=1$ and varying $\rho_m$ are illustrated in Figure~\ref{stabilityKidadensity}. Similarly, results for vortices with $\alpha=\{0.25,\,0.5\}$ and non-uniform density excess  are illustrated in Figure~\ref{stabilitynonKidadensity}. For these calculations, $\chi$ was varied by changing $\omega_m$ while $\rho_m$ was chosen such that a fixed mass per unit length was added to the vortex with specified $\alpha$ and $\beta$ for all $\chi$. As the vortex core boundaries all pass through $(0,1)$, central density increases with $\chi$. The central density for the vortices illustrated in Figures~\ref{stabilityKidadensity} and \ref{stabilitynonKidadensity} are shown as a function of $\chi$ in Figure~\ref{densitycurves}. Central densities range from $3$ to $12$ times the background level at large values of $\chi$. 

	In Figure~\ref{stabilityKidadensity} we can see the effect of introducing a small density excess on the stability of a Kida vortex. For the case with $\rho_m=0.5$, results are shown for streamlines with $y_{max}=0.5$ and $0.85$. In the former case where departures from a quadratic stream function are smaller, a parametric instability band is seen emanating from the expected location $\chi=4.65$. This appears to attempt to connect with the band originating from large $\chi$ and $\theta$ in the Kida vortex case, leaving a region that is very weakly unstable, or possibly stable, between them. Bands with such regions seem to be a common feature in these calculations. They require time consuming calculations at high resolution to locate them. In our case we found it impractical to resolve instabilities with growth rates $< 0.0001\Omega_P$. 

	When the streamline with $y_{max}=0.85$ is considered, the band originating from $\chi \sim 4.65$ has broadened to produce an unstable region with growth rate $\sim 0.05\Omega_P$ for $4< \chi < 5$. Two additional narrow bands appear at larger $\chi$ with characteristic growth rates $\sim 0.01\Omega_P$. When the ensemble of streamlines is considered for this model, it is difficult to find any $\chi$ for which there is stability for all $\theta$. The parameter $\omega_m$ is varied to produce the range of aspect ratios shown. The lower panels of Figure~\ref{stabilityKidadensity} illustrate the stability of a model with $\rho_m=2.0$. This higher density model shows qualitatively similar behaviour but with the appearance of a larger number of weak instability bands and,  for $y_{max}=0.85,$  an instability band for $\chi<4$ that is similar to that seen in the models with centrally concentrated Bernoulli source and no density excess. 
	  
	The results for the models with $\alpha=0.25$ and $\alpha=0.5$ illustrated in Figure~\ref{stabilitynonKidadensity} are for streamlines that pass through $(0.0,0.85)$. The behaviour is qualitatively similar to the previous cases with density excesses,  but with growth rates at large $\chi$, $\sim 0.05\Omega_P$ being similar to the concentrated Bernoulli source cases without density excesses.

\begin{figure*}
		\begin{minipage}{177mm}
			\begin{center}
				\subfigure[Streamline at $(0,0.95)$.]
				{
				\includegraphics[width=0.45\textwidth]{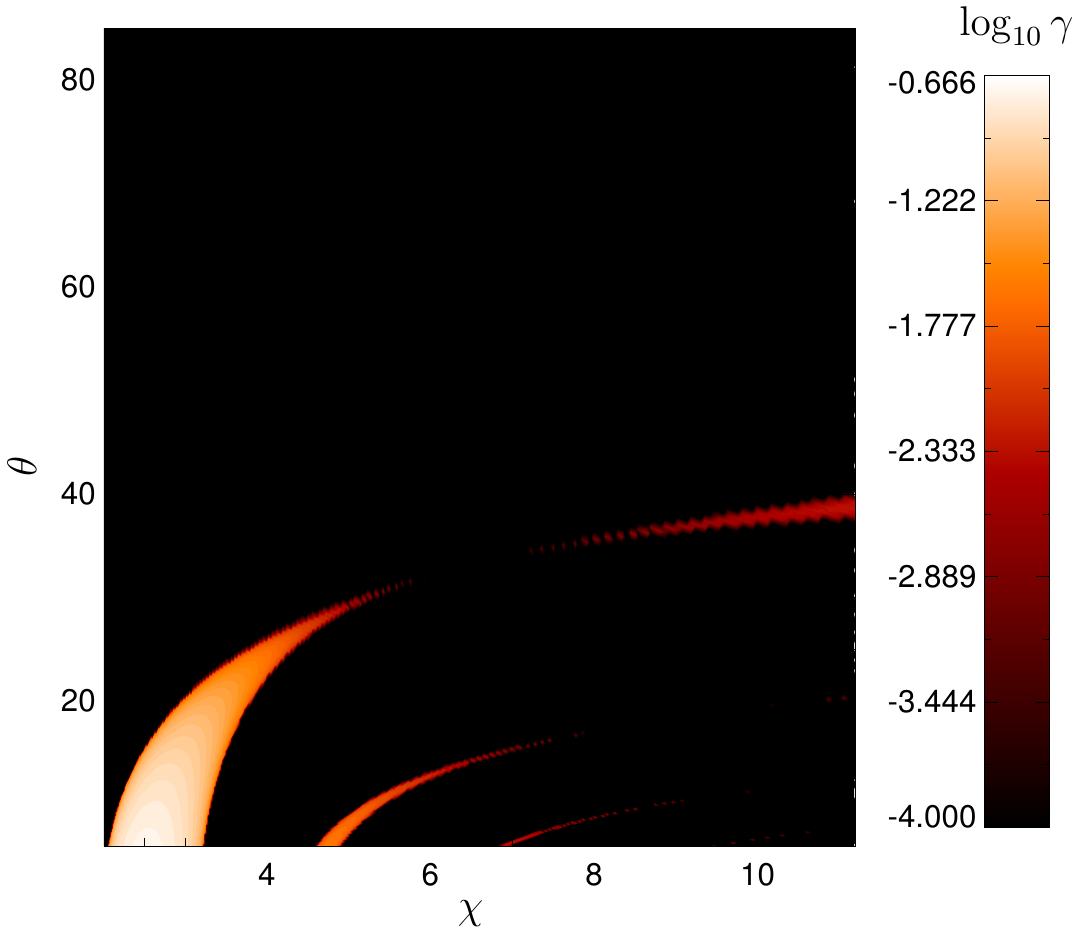}
				}
				\subfigure[Streamline at $(0,0.95)$.]
				{
				\includegraphics[width=0.45\textwidth]{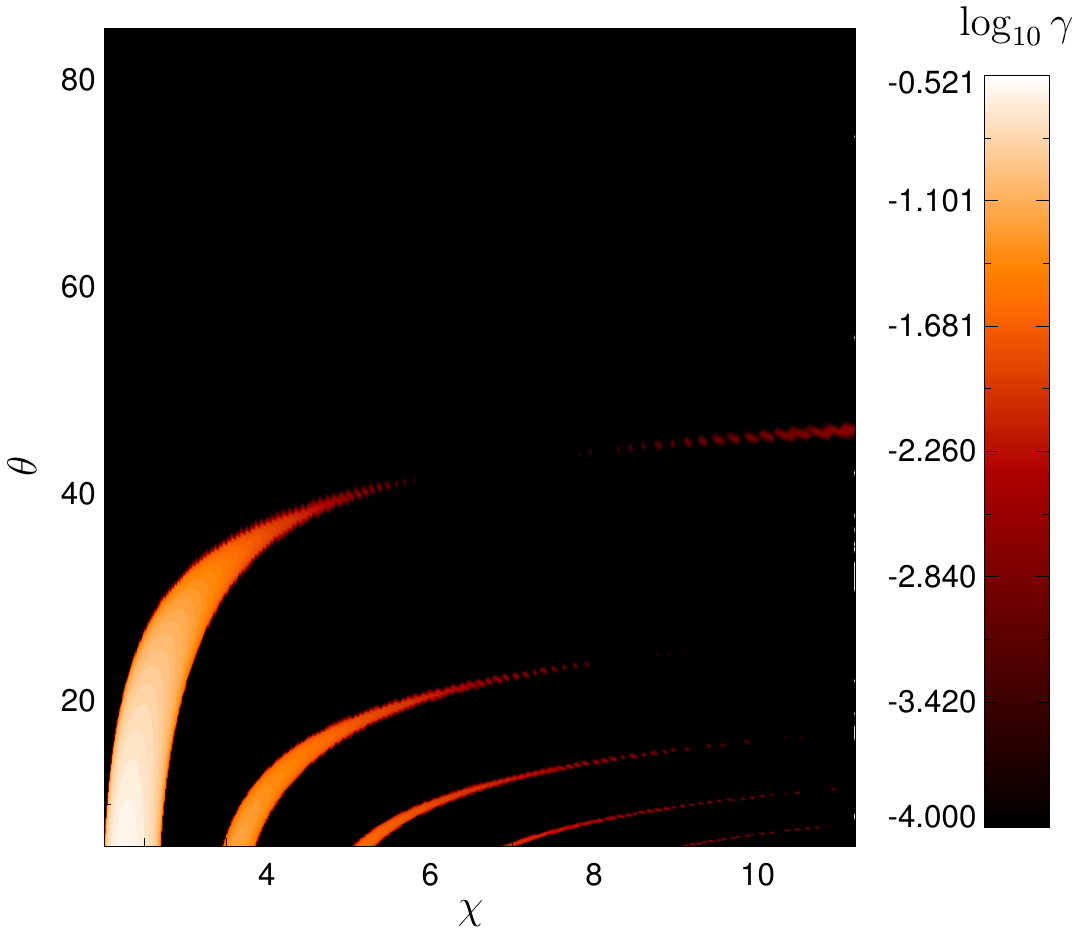}
				}
			\caption{The stability of the polytropic models in a non-Keplerian background. Cases with maximum density equal to  three times the background value a) and six times the background value b).}
			\label{stabilitypolytrope}
			\end{center}
		\end{minipage}
	\end{figure*}

	In Figure~\ref{stabilitypolytrope} we show the stability of the polytropic models with $n_1=1$ in the $(\chi,\theta)$ plane with central densities $3$ and $5$ times the background level. The results are for streamlines close to the core boundary with $y_{max}=0.95$. We recall that for these models, vortices associated with values of $\chi \ne 7$, have to be considered to be immersed in a non-Keplerian background (see Section~\ref{polyeq} ). The presence of instability bands will be noted, three in the lower central density case and four in the higher central density case. In this respect the results are qualitatively similar to models with central density excess immersed in a Keplerian background for all $\chi$. The characteristic growth rates for $\chi \sim 7$ are $\sim 0.01\Omega_P$.

\begin{figure}
			\includegraphics[width=0.45\textwidth]{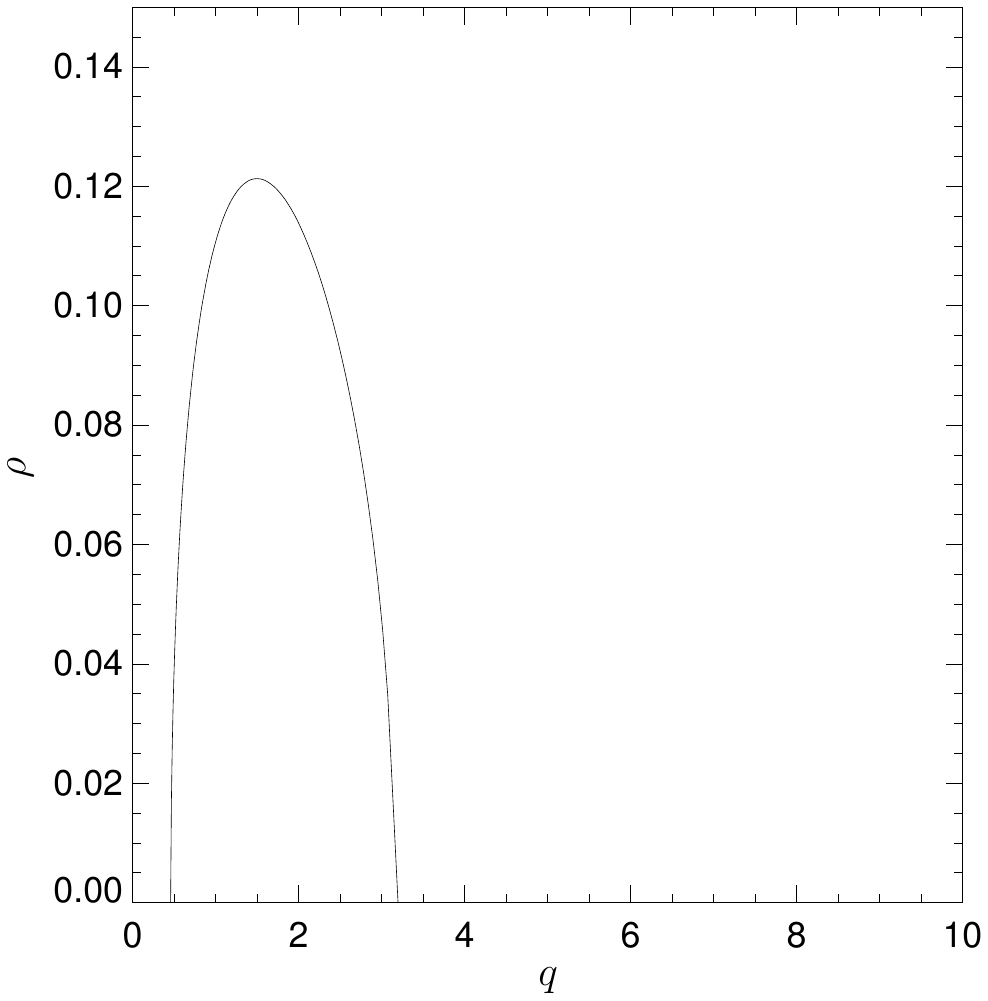}
			\caption[]{The growth rate of the parametric instability of the polytropic model as a function of $q$ expressed in units of $\Omega_P^2$ for $k_z=0$.}
			\label{parametriccurve}
     \end{figure}

	Finally, we have studied the parametric instability that can occur when the vertical stability of the polytropic model is considered (see Section \ref{vertparam}) because that has no internal shear. This connects to a Keplerian background for $\chi=7$. Accordingly we consider that case. The growth rate is plotted as a function of $q/\Omega_P^2$ in Figure~\ref{parametriccurve}. It will be seen that there is instability for $q\gtrsim 0.5\Omega_P^2$ and the growth rate reaches a maximum of $\sim 0.12\Omega_P$ for $q=1.5\Omega_P^2$. This is about an order of magnitude larger than the growth rates illustrated in Figure~\ref{stabilitypolytrope} for modes with time-independent wavenumber at this value of $\chi$. Thus modes with $k_z=0$ \citep{Chang2010} may dominate in this spacial case. However, as  noted in section \ref{timeind},  the neglect of vertical stratification is likely to restrict validity to regions close to the midplane. The maximum value of $q$ for the polytropic models occurs on the core boundary and is $q_{max}= 7n_1b\Omega_P^2/4$. Thus stability is indicated for sufficiently small $n_1 b < \sim 2/7$.
  
\section {Discussion and Conclusions}\label{sec:conclusions}
	In this paper we have generalised the steady state, two dimensional Kida solution in the shearing sheet to allow for general vorticity profiles and dust density distributions. The length scales are assumed to be small enough that the gas flow can be approximated as being incompressible. The solutions apply in the limit of very small dust stopping time so that the dust density distribution is frozen-in such that the system can be modelled as an incompressible fluid with variable density. As this can be specified arbitrarily, a wide range of solutions is in principle possible. In particular, we generalised the well known Kida solution so that it can incorporate a central dust density enhancement (see the polytropic models of Section \ref{polyeq}). However, the background disc is Keplerian only for an aspect ratio $\chi=7$. Other values of $\chi$ correspond to backgrounds with density maxima or minima. Other solutions obtained numerically were discussed in Section \ref{sec:numerical_results}. 

	These solutions, although two dimensional, can apply to horizontal planes in an isothermal disc while being in hydrostatic equilibrium in the vertical direction. It is important to note that the Kida solution is special in that the velocity inside the core is a linear function of coordinates that results in in a constant period for circulating around streamlines, resulting in zero internal shear. This is not the case outside the core or inside it for general vorticity distributions, with significant consequences for a discussion of stability. 

	We have generalised the stability analysis applied to the core of the Kida vortex so that it can be extended to more general models in Section \ref{stability}. 
The core of a Kida vortex is special because the stability problem becomes separable for a specific choice of time-dependent wavenumber. This does not apply in general.
 However, we found that it is possible to look for modes localised on individual streamlines and that, from an Eulerian viewpoint, these can have either a time-independent wavenumber or a wavenumber that depends on time.
 When the circulation period is not constant on streamlines, the latter form ultimately increases linearly with time and so cannot be associated with conventional 
exponentially growing linear modes (see Section \ref{kpropt}), a situation familiar in shearing box calculations. 

	The modes with time-independent wavenumber can be exponentially growing and turn out to yield those found in the Kida vortex case. For only these can the analysis can be extended to apply to more general vortices. We remark that when comparing the situation to that of perturbations of an axisymmetric disc, the time-independent modes correspond to axisymmetric $(m=0)$ modes, while the time-dependent wavenumber modes correspond to $m\ne0$. 

	  Following our procedure we are able to generalise the strong instability of Kida vortices for small aspect ratios, $\chi < 4$ (associated with a central saddle point in the pressure distribution) to apply to the centre of vortices in general. Apart from having a strong instability, vortices of this type are of less interest as they will not attract dust which tends to migrate towards pressure maxima. We also showed  how parametric instability bands should appear moving outwards from the centre. This behaviour was found to be consistent with our numerical results presented in Section \ref{satbcalc}.

	We find that vortices with a concentrated vorticity source and no density excess have strong instability bands at all aspect ratios of interest with growth rates $\sim 0.05\Omega_P$. This is seen for the limiting case of a point vortex and we can infer that this is a generic issue when considering dust spiralling inwards from the outer disc towards the vortex core. Models with a density excess can show many narrow parametric instability bands though those with flatter vorticity profiles show less strongly growing modes with growth rates $\sim 0.01\Omega_P$.

	We also considered the stability of the dust laden polytropic model with $\chi =7$ to local modes with time-dependent wavenumber and with $k_z =0$. These are effectively those considered by \citet{Chang2010}. This is possible as in this special case the vortex has no internal shear. Indeed we found that modes could occur with growth rates, $\gamma,$ up to $0.1\Omega_P$. However, these modes would be affected by shear if an attempt is made to generalise them to other cases. Even cases with weak shear shown in Figure~\ref{equilKidadensity} correspond to shear $\sim 0.01\Omega_P$. For modes with putative growth rates $\sim 0.1\Omega_P$ we might accordingly expect temporary growth for $10$ growth times or a temporary amplification factor $\sim 10^4$. We believe that a non-linear analysis is required to resolve the outcome in such a case. Note too as remarked in section \ref{timeind} these modes are also likely to be affected
by vertical stratification unless streamlines very close to the midplane are considered.

	However, the results presented here taken together imply that dust particles attracted from the outer disc to a vortex core with high aspect ratio $\chi$ may well encounter parametric instabilities with characteristic growth rates of a few $\times 10^{-2}\Omega_P$ up to a maximum of $0.1\Omega_P$. This is the case even outside any high density core and so it is important to assess potential consequences for dust accumulation in vortices. The instabilities are parametric and local so that they can be inhibited by either dissipative effects or effects that disrupt the periodicity of the circulating motion. As the magnitude of dissipative effects is uncertain and for example dependent on the assumed location in a protoplanetary disc, we shall consider the second type of effect.

	In this context we note that the accumulation of dust in vortices may occur rapidly \citep[eg.][]{Lyra2009, Meheut2012} such that the dust particle motion departs significantly from being periodic. This may be the case for the gas motion also \citep{Meheut2012}. For these reasons, parametric instability may not have been seen in simulations up to now. However, if we suppose parametric instability is present, it is likely to lead to some form of low level turbulence,  This was indicated by work of both  \cite{Lesur2010} and \citet{Lyra2011} who found that such instabilities do not have a strongly disruptive   effect on large aspect ratio vortices produced by the subcritical baroclinic instability (SBI). Thus although parametric instability may  act to cause a vortex to  ultimately  decay, it  may  be maintained
 if there is some mechanism to generate it, such as the SBI or Rossby wave instability. 
For Kida vortices,  strong, exponentially growing and potentially  fast vortex-destroying instability only occurs on account of the saddle point instability which for
  which $3/2 \lesssim \chi \lesssim 4$.

	For larger aspect ratios we might expect that there is a balance between inward flow due to the mean pressure gradient and turbulent diffusion \citep[eg.][]{Lyra2013}. To make a crude estimate this we note that the inflow rate for small particles driven by the pressure gradient is $|{\bf v}| \sim |\nabla P|/\rho \tau_s$ \citep[eg.][]{Papaloizou2006}.

	Supposing the vortex has a length scale $L$ in the minor axis direction, we estimate $P \sim \rho \Omega_P^2 L^2$ and $|{\bf v}| \sim \Omega_P^2L\tau_s$. As the unstable modes are local, the wavelength should be $\ll L$. For the purpose of making crude estimates we adopt $\pi/|{\bf k}|=L/10$. An estimate of the associated diffusion coefficient based on dimensional scaling is $D=\gamma/|{\bf k}|^2$. Balancing pressure driven inflow against diffusion we obtain $|\nabla \rho|/|\rho| = 1/L_{\rho} \sim |{\bf v}|/D\sim (10\pi \Omega_P)^2 \tau_s./(\gamma L)$. Hence $L_{\rho}\sim fL,$ where $f \sim (\gamma/\Omega_P)/(100\pi^2 \Omega_P\tau_s)$. Significant dust concentrations become possible once $f<1$. For $\gamma=0.1\Omega_P$, this becomes equivalent to $\Omega_P\tau_s > \sim 10^{-4}$. Note that the characteristic inflow time is then $1/(\Omega_P^2\tau_s)$. Although estimates of the above type are highly uncertain they 
are made to indicate that the existence of parametric instabilities of the type considered here does not necessarily prevent the possibility of dust accumulation in vortices.
\section*{Acknowledgements}
The authors would like to thank the anonymous reviewer for their helpful and insightful comments which lead to numerous improvements in the manuscript. ADR acknowledges support by STFC.
 
\bibliography{references}

\end{document}